\documentclass[utf8]{FrontiersinHarvard} 

\usepackage{url,hyperref,lineno,microtype,subcaption}
\usepackage[onehalfspacing]{setspace}
\usepackage{caption}
\usepackage{graphicx}
\usepackage{threeparttable}
\usepackage{comment}
\usepackage{array}

\newcommand\chandra{{\it Chandra}}

\newcommand\xmm{{\it XMM-Newton}}

\newcommand\hexp{{\it HEX-P}}
\newcommand\integral{{\it INTEGRAL}/IBIS}

\newcommand\swift{{\it Swift\/}}
\newcommand\nustar{{\it NuSTAR}}
\newcommand\suzaku{{\it Suzaku}}
\newcommand\erosita{{\it eROSITA}}
\newcommand\gaia{{\it Gaia}}

\def\amin{\ifmmode^{\prime}\else$^{\prime}$\fi}
\def\asec{\ifmmode^{\prime\prime}\else$^{\prime\prime}$\fi}
\def\simgt{\lower.5ex\hbox{$\; \buildrel > \over \sim \;$}}
\def\simlt{\lower.5ex\hbox{$\; \buildrel < \over \sim \;$}}
\newcommand{\sgra}{\ensuremath{\text{Sgr\,A}^{\star} }}

\def\keyFont{\fontsize{8}{11}\helveticabold }
\def\firstAuthorLast{Mori {et~al.}} 
\def\Authors{Kaya Mori\,$^{1,*}$, Gabriele Ponti\,$^{2,3}$,  Matteo Bachetti\,$^{4}$, Arash Bodaghee\,$^{5}$, Jonathan Grindlay\,$^{6}$, Jaesub Hong\,$^{6}$,  Roman Krivonos\,$^{7}$, Ekaterina Kuznetsova\,$^{7}$,  Shifra Mandel\,$^{1}$, Antonio Rodriguez\,$^{8}$, Giovanni Stel\,$^{2}$, Shuo Zhang\,$^{9}$,  Tong Bao\,$^{10}$, Franz Bauer\,$^{11}$, Ma\"ica Clavel\,$^{12}$, Benjamin Coughenour\,$^{13}$,  Javier A. Garc\'ia\,$^{14}$, Julian Gerber\,$^{1}$,   Brian Grefenstette\,$^{8}$, Amruta Jaodand\,$^{8}$, 
 Bret Lehmer\,$^{15}$,  Kristin Madsen\,$^{14}$, Melania Nynka\,$^{16}$, Peter Predehl\,$^{3}$, Ciro Salcedo\,$^{1}$,  Daniel Stern\,$^{17}$ and John Tomsick\,$^{13}$}
\def\Address{\footnotesize $^{1}$ Columbia Astrophysics Laboratory, Columbia University, New York, NY, USA \\
$^{2}$ INAF - Osservatorio Astronomico di Brera, Merate, Italy \\ 
$^{3}$ Max-Planck-Institut für extraterrestrische Physik, Garching, Germany \\
$^{4}$ INAF-Osservatorio Astronomico di Cagliari, Selargius, Italy\\ 
$^{5}$ Dept. of Chemistry, Physics and Astronomy, Georgia College $\&$ State University, Milledgeville, GA, USA \\ 
$^{6}$ Harvard-Smithsonian Center for Astrophysics, Harvard University, Cambridge, MA, USA \\ 
$^{7}$ Space Research Institute (IKI), Russian Academy of Sciences, Moscow, Russia\\ 
$^{8}$ Cahill Center for Astronomy and Astrophysics, California Institute of Technology,  Pasadena, CA, USA\\
$^{9}$ Department of Physics and Astronomy, Michigan State University, East Lansing, MI, USA \\ 
$^{10}$ School of Astronomy and Space Science, Nanjing University, Nanjing, China \\ 
$^{11}$ Instituto de Astrof{\'{\i}}sica, Pontificia Universidad Cat{\'{o}}lica de Chile, Santiago, Chile; Millennium Institute of Astrophysics, Santiago, Chile\\ 
$^{12}$ Univ. Grenoble Alpes, CNRS, IPAG, 38000 Grenoble, France \\ 
$^{13}$ Space Sciences Laboratory, UC Berkeley, Berkeley, CA, USA\\ 
$^{14}$ X-ray Astrophysics Laboratory, NASA Goddard Space Flight Center, Greenbelt, MD, USA \\ 
$^{15}$ Department of Physics, University of Arkansas, Fayetteville, AR, USA \\ 
$^{16}$ Kavli Institute For Astrophysics and Space Research, Massachusetts Institute of Technology, Cambridge, MA, USA \\ 
$^{17}$ Jet Propulsion Laboratory, California Institute of Technology, Pasadena, CA, USA
}

\def\corrAuthor{\footnotesize Kaya Mori}

\def\corrEmail{\footnotesize kaya@astro.columbia.edu}

\begin{document}
\onecolumn
\firstpage{1}

\title {The High Energy X-ray Probe (HEX-P): resolving the nature of Sgr A* flares, compact object binaries and diffuse X-ray emission in the Galactic Center and beyond} 

\author[\firstAuthorLast ]{\Authors} 
\address{}
\correspondance{} 

\extraAuth{}

\maketitle

\begin{abstract}

\emph{HEX-P} is a probe-class mission concept that will combine high spatial resolution X-ray imaging ($<10''$ FWHM) and broad spectral coverage (0.2--80 keV) with an effective area far superior to current facilities' (including \emph{XMM-Newton} and \emph{NuSTAR}). These capabilities will enable revolutionary new insights into a variety of important astrophysical problems. We present scientific objectives and simulations of \emph{HEX-P} observations of the Galactic Center (GC) and Bulge. We  demonstrate the unique and powerful capabilities of the \emph{HEX-P} observatory for studying both X-ray point sources and diffuse X-ray emission. 
\emph{HEX-P} will be uniquely equipped to explore a variety of major topics in Galactic astrophysics, allowing us to (1) investigate broad-band properties of X-ray flares emitted from the supermassive black hole (BH) at Sgr A* and probe the associated particle acceleration and emission mechanisms; (2) identify hard X-ray sources detected by \emph{NuSTAR} and determine X-ray point source populations in different regions and luminosity ranges; (3) determine the distribution of compact object binaries in the nuclear star cluster and the composition of the Galactic Ridge X-ray emission; (4) identify X-ray transients and measure fundamental parameters such as BH spin; (5) find  hidden pulsars in the GC; (6) search for BH--OB binaries and hard X-ray flares from young stellar objects in young massive  clusters; (7) measure white dwarf (WD) masses of magnetic CVs to deepen our understanding of CV evolution and the origin of WD magnetic fields;  (8) explore primary particle  accelerators in the GC in synergy with future TeV and neutrino observatories; (9) map out cosmic-ray distributions by observing non-thermal X-ray filaments; (10) explore past X-ray outbursts from Sgr A* through X-ray reflection components from giant molecular clouds.  

\tiny
 \keyFont{ \section{Keywords:} Galaxy: centre, black hole physics, X-rays: binaries, stars: cataclysmic variables, accretion, acceleration of particles, pulsars: general} 
\end{abstract}

\section{Introduction}

The Galactic Center (GC) harbors an extremely dense and diverse population of stars, compact objects, X-ray binaries (XRBs), molecular clouds, magnetic filaments, and energetic cosmic-ray accelerators within its central few degrees.   
Sgr A*, the radiative counterpart of the supermassive black hole (SMBH), plays a crucial role in the gas dynamics and binary formation processes in the central few parsecs, where stars and XRBs cluster \citep{Muno2003, Mori2021}. Despite its faint quiescent X-ray emission, Sgr A* intermittently accelerates particles and emits daily X-ray flares. In the past, Sgr A* is believed to have been 
far more active, possibly fueling the Fermi GeV bubbles \citep{Cheng2011}, X-ray chimneys \citep{Ponti2021}, and, as recently as a few hundred years ago, X-ray emission from the GC molecular clouds \citep{Ponti2010, Clavel2013}. The population of relativistic cosmic rays, including leptons and hadrons, is evident in radio/X-ray filaments and TeV emission from molecular clouds \citep{Zhang2014, Zhang2020}. 

X-ray observations provide valuable insights into compact objects, their binary systems, and particle accelerators in the GC region, since those sources are frequently observed to emit X-rays. However, conducting X-ray studies in this densely populated region, which is located at a large distance of 8 kpc, poses challenges. 
Furthermore, the GC is filled with numerous point and diffuse X-ray sources, requiring high-angular-resolution X-ray telescopes to resolve them. 
Over the past two decades, extensive surveys of the GC and bulge regions in the soft X-ray band ($< 10$ keV) have been carried out by \chandra, \xmm\ and \suzaku. In a $2^\circ\times0.8^\circ$ region, \chandra\ detected $\sim 10,000$ X-ray point sources below 8 keV \citep{Wang2002, Muno2009, Zhu2018}. Studies of diffuse X-ray emission in the GC and Bulge regions have been conducted by \suzaku\ and \xmm, revealing extensive hot plasma emission \citep{Koyama2009, Ponti2015b, Anastasopoulou2023}. Frequent X-ray monitoring of the GC with \swift-XRT resulted in the detection of $\sim20$ X-ray transients in the GC \citep{Degenaar2015}. Follow-up X-ray observations with \chandra, \xmm\ and \nustar\ enabled the identification of some of those transients \citep{Mori2013, Mori2019}. 

In the GC and bulge regions, soft X-ray data 
is limited to $E>2$ keV due to significant X-ray absorption and scattering. Consequently, the narrow-band (2--10 keV) X-ray data obtained by \chandra\ and \xmm\ are often insufficient for identifying 
source types and accurately measuring 
spectral parameters (e.g., plasma temperatures and photon indices) due to the parameter degeneracy with neutral hydrogen column density ($N_{\rm H}$). In the GC, $N_{\rm H}$ is typically high ($N_{\rm H} \sim(5-20)\times10^{22}$~cm$^{-2}$) and variable across different regions \citep{Johnson2009}, complicating our ability to constrain other spectral parameters in the soft X-ray band. 
In the higher energy band above 10 keV, \integral\ and \suzaku-HXD unveiled diffuse hard X-ray emission and a handful of bright XRBs \citep{Revnivtsev2004, Yuasa2008}. Subsequently, \nustar\ surveyed $\sim2/3$ of the central $2^\circ\times0.8^\circ$ region, resolving the hard X-ray emission better with its sub-arcminute angular resolution. However, the 
\nustar\ GC observations 
are significantly affected by contamination from stray-light and ghost-ray background photons. As a result, several GC and bulge regions, including the Limiting Window (one of the low extinction windows 
where \chandra\ detected hundreds of soft X-ray point sources), Sgr C (one of the GC molecular clouds), and the southern part of the central $2^\circ\times0.8^\circ$ region, remain unobserved by \nustar\ due to the elevated background levels. 
While \chandra, \xmm\ and \suzaku\ have provided a comprehensive \emph{soft} X-ray view of the GC region, our \emph{broad-band} knowledge of the X-ray point and diffuse sources in the GC remains relatively limited. 

In this paper, we highlight the unique capabilities of \hexp\ to significantly advance our understanding of X-ray-emitting objects within the GC, Bulge and other regions in our Galaxy. Specifically, \hexp's exceptional angular resolution and broad-band X-ray coverage extending up to 80 keV are crucial for resolving  X-ray sources, characterizing their X-ray emission, and identifying their nature. Given its energy band -- extending to 25 keV with $3.5''$ angular resolution, LET will be highly advantageous for investigating X-ray sources in crowded fields like the GC. On the other hand, HET will allow us to investigate hard X-ray sources and non-thermal X-ray emission with far better sensitivity than \nustar. X-ray spectral and timing properties of transients will be fully characterized by \hexp\ thanks to its fast readout time. Both LET and HET will provide excellent Fe line diagnostics at $E=6\rm{-}7$ keV; detecting and resolving highly-ionized Fe lines at 6.7 and 6.9 keV will enable us to discern between thermal and non-thermal X-ray sources. Moreover, a neutral Fe fluorescence line at 6.4 keV can be utilized to  constrain X-ray reflection models for molecular clouds and magnetic CVs (mCVs). Additionally, both LET and HET possess high-resolution timing capabilities, making them valuable tools for identifying pulsars and millisecond pulsars (Alford et al. 2023;  Jaodand et al. 2023). During the 2030s, \hexp\ is poised to make unique contributions to multi-messenger astrophysics of the GC, Bulge and Plane, in synergy with existing and upcoming telescopes across various wavelengths. These include the CTAO for TeV gamma-ray observations, the IceCube gen2 neutrino experiment, radio telescopes such as EHT (Event Horizon telescope), SKA (Square Kilometer Array) and next-generation VLA, as well as infrared telescopes like JWST and GRAVITY. 

The paper is organized as follows. \S\ref{sec:hexp_design} describes the current design of the \hexp\ telescope and simulation tools. \S\ref{sec:gc_program} briefly introduces key science cases in the primary GC observation program. More detailed descriptions can be found about Sgr A* flares (\S\ref{sec:sgra_flare}), X-ray source populations in the GC and Bulge (\S\ref{sec:gc_pointsources} and diffuse X-ray sources in the GC (\S\ref{sec:diffuse_sources}). \S\ref{sec:star_forming} and \S\ref{sec:mcv} demonstrate some examples for potential \hexp\ GO program observations of star-forming regions and CVs. \S\ref{sec:conclusion} concludes the paper.  

\section{\emph{HEX-P} mission design and simulation \label{sec:hexp_design}}

The \textit{High Energy X-ray Probe} (\hexp; Madsen et al.\ 2023) is a probe-class mission concept that offers sensitive broad-band X-ray coverage (0.2--80\,keV) with exceptional spectral, timing and angular capabilities. It features two high-energy telescopes (HETs) that focus hard X-rays and one low-energy telescope (LET) that focuses lower-energy X-rays.

The LET consists of a segmented mirror assembly coated with Ir on monocrystalline silicon that achieves an angular resolution of $3.5''$, and a low-energy DEPFET detector, of the same type as the Wide Field Imager (WFI; \citealp{Meidinger2020}) onboard \textit{Athena} \citep{Nandra2013}. It has $512 \times 512$ pixels that cover a field of view of $11.3'\times 11.3'$. The LET has an effective passband of 0.2--25\,keV, and a full frame readout time of 2\,ms, which can be operated in a 128 and 64 channel window mode for higher count rates to mitigate pile-up with faster readout. Pile-up effects remain below an acceptable limit of ${\sim}1\%$ for fluxes up to ${\sim}100$\,mCrab in the smallest window configuration. Excising the core of the PSF, a common practice in X-ray astronomy, will allow for observations of brighter sources, with a typical loss of up to ${\sim}60\%$ of the total photon counts.

The HET consists of two co-aligned telescopes and detector modules. The optics are made of Ni-electroformed full shell mirror substrates, leveraging the heritage of \textit{XMM-Newton}, and coated with Pt/C and W/Si multilayers for an effective passband of 2--80\,keV. The high-energy detectors are of the same type as those flown on \nustar, and they consist of 16 CZT sensors per focal plane, tiled $4\times4$, for a total of $128 \times 128$ pixel spanning a field of view of $13.4' \times 13.4'$.
The HET utilizes the same optics technology as \xmm\ 
and the PSF is energy-dependent with an
HPD of $10''$ at 3 keV, $\sim17''$ at 20 keV, and increases
at higher energies. For the purpose of simulations in
this paper, an average HPD of $17''$ was used across
the entire bandpass. 

All simulations presented here were produced with a set of response files that represent the observatory performance based on current best estimates (see Madsen+23). The effective area is derived from a ray-trace of the mirror design including obscuration by all known structures. The detector responses are based on simulations performed by the respective hardware groups, with an optical blocking filter for the LET and a Be window and thermal insulation for the HET. The LET background was derived from a GEANT4 simulation \citep{Eraerds2021} of the WFI instrument, and the one for the HET from a GEANT4 simulation of the \nustar\  instrument, both positioned at L1. 
Throughout the paper, we present our simulation results for \hexp\ using the SIXTE \citep{Dauser2019} and XSPEC toolkits (version 12.13.0;  \citep{Arnaud1996}). To ensure the most realistic simulation results, we incorporated recent high-resolution X-ray images (mostly from \chandra\ or other wavelength observations), the best-known spectral information, and theoretical model predictions. Various exposure times have been considered for the feasibility studies presented in the following sections.

\section{\emph{HEX-P}'s Galactic Center observation program \label{sec:gc_program}}

\begin{figure}[h!]
\begin{center}
\includegraphics[width=1.0\textwidth]{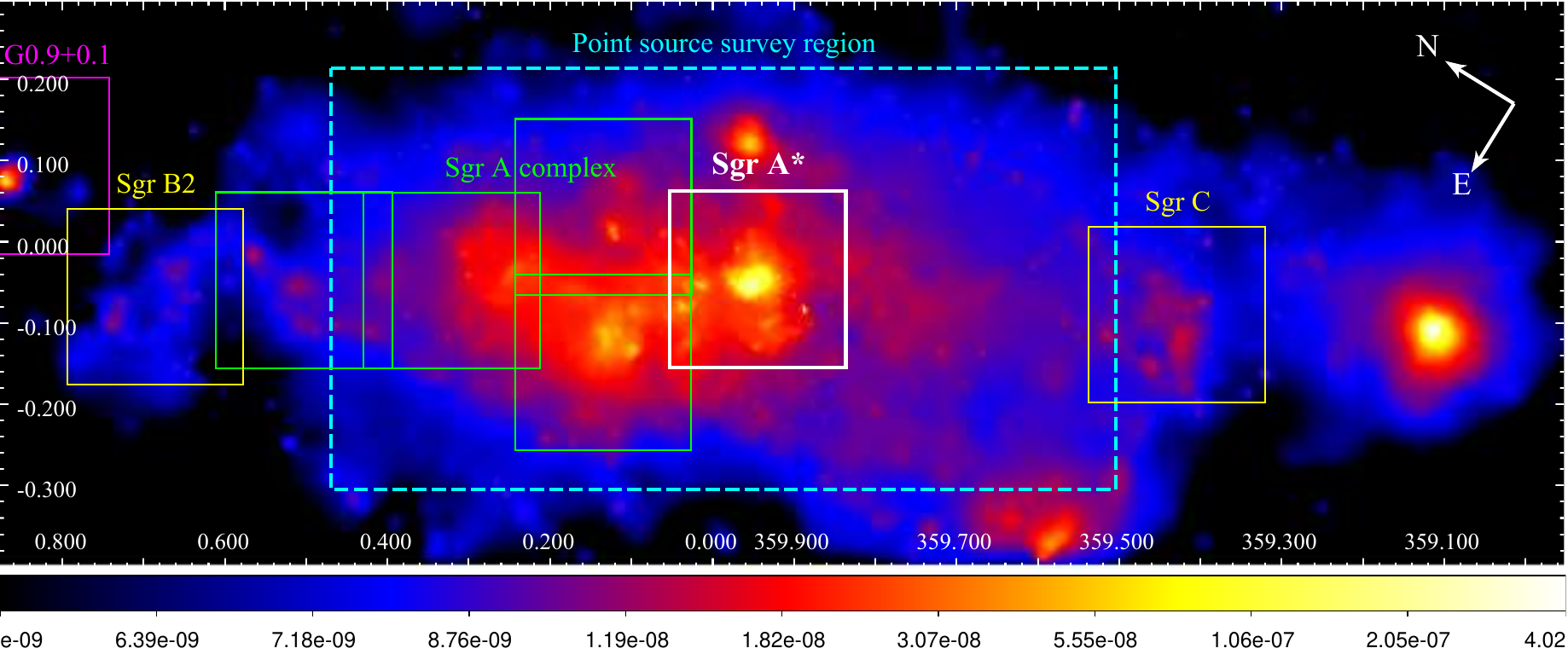}
\includegraphics[width=1.0\textwidth]{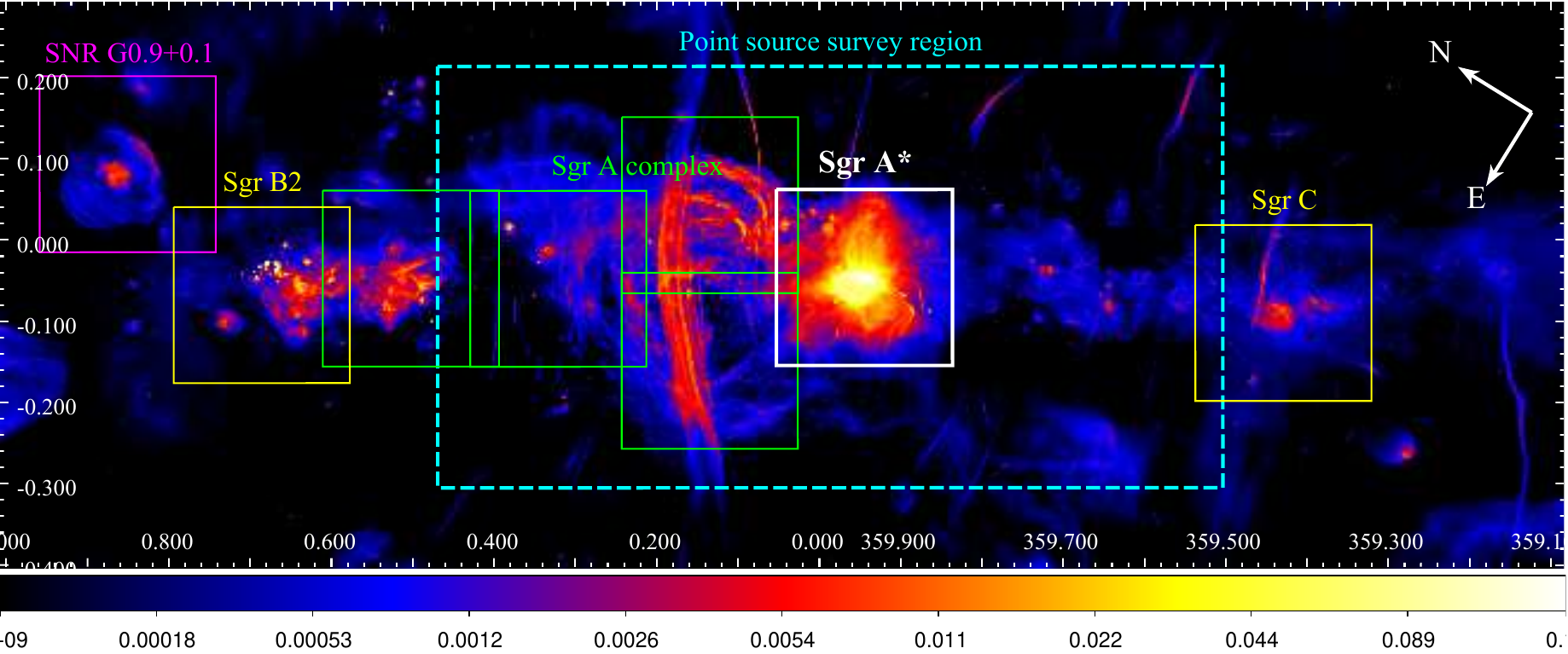}
\end{center}
\caption{The GC regions surveyed by the \hexp\ primary observation program, overlaid with \chandra\ X-ray (top) and MeerKAT radio images (bottom). SNR G1.9+0.3 and the LW are located outside the images above. 
\chandra\ image credit: 
\citet{Wang2002, Muno2009}. MeerKAT image credit: \citet{Heywood2022}.} \label{fig:obs_strategy}
\end{figure}

Throughout the paper, we present how \hexp\ will have a high impact on exploring and understanding a variety of high-energy sources in the GC. With its versatile and well-balanced capabilities, including (1)  large effective area in 0.2--80 keV, (2) high energy resolution for resolving atomic lines, (3) low background levels, (4) $<3.5''$ (LET) and $<20''$ (HET) angular resolutions, and (5) $<2$ ms timing resolution, \hexp\ is an ideal X-ray probe mission for investigating a diverse class of X-ray point and diffuse sources. We have outlined the targets for \hexp's primary science program  in Table \ref{tab:obs} and briefly described them below. Figure \ref{fig:obs_strategy} displays the survey regions covered by the primary science program, along with \chandra\ X-ray and MeerKAT radio images of the GC region.

\begin{table*}[h!]
\renewcommand\thetable{1}
{\small
\begin{center}
\caption{\hexp's primary observation program of the Galactic Center region}
\begin{tabular}{lccc}
\hline\hline
Target name & Exposure time (ks) & Science objectives \\ 
\hline\hline
Sgr A* and the central 10 pc & 500 & Sgr A* flares, NSC 
 and filaments \\
Sgr A complex   & 300 (with 4 tiles) & Molecular clouds, filaments, star clusters, point sources \\
Sgr B2  & 100 &  Molecular clouds and point sources \\
Sgr C & 100 & Molecular clouds and point sources \\ 
Limiting Window   & 300 &  Point sources \\
$1^\circ\times0.7^\circ$ survey &  850 (17 tiles $\times$ 50 ks) & Point sources \\ 
SNR G1.9+0.3 & 100 & Youngest SNR in our Galaxy \\ 
PWN G0.9+0.1 & 50 & One of the brightest TeV sources \\ 
\hline
\label{tab:obs}
\end{tabular}
\end{center}
Note: 2.3 Ms total exposure is currently allocated to the \hexp's primary science observations of the GC regions as listed above.  }
\end{table*}

\textbf{(1) Broad-band X-ray investigation of Sgr A* X-ray flares (\S \ref{sec:sgra_flare}):} \hexp\ observations of Sgr A* X-ray flares will play a unique and impactful role in the multi-messenger astrophysics of studying how the supermassive BH accelerates particles and emit flares in various ways. \hexp\ will fully characterize the spectral and timing properties of bright flares thanks to the  broadband and continuous coverage without earth occultation. In the 2030s, \hexp\ in synergy with other telescopes in the radio (EHT), IR (GRAVITY) and TeV (CTA) bands will provide a wealth of multi-wavelength SED and lightcurve data on Sgr A* flares. These legacy data sets will make a significant impact on understanding the particle acceleration and emission mechanisms of Sgr A* flares.

\textbf{(2) Revealing X-ray point source populations in the GC (\S\ref{sec:source_population}, \ref{sec:transients}, \ref{sec:pulsars}): } The broadband X-ray spectral and timing capabilities of \hexp\ will provide a variety of useful diagnostic tools for identifying X-ray point sources and transients in the GC, Bulge, and beyond. These diagnostic tools encompass Fe line analysis, broad-band X-ray spectroscopy, pulsation detection, and X-ray variability study. In the primary science program, \hexp\ is expected to detect $\sim1,000$ X-ray sources above 8 keV and identify/classify them with existing or future IR and X-ray observation data. 
 In particular, \hexp\ will probe the X-ray luminosity range of $L_X \sim 10^{32}\rm{-}10^{33}$ erg\,s$^{-1}$, poorly explored by \nustar\ due to the limited sensitivity. This is the regime where a large number of the \chandra\ sources, mostly composed of compact object binaries, remain unclassified. As demonstrated by \nustar, \hexp\ will obtain broadband X-ray spectral and timing data of X-ray transients, will be instrumental in detecting pulsations (in the case of transient magnetars) and BH spins (in the case of BH transients) as well as characterizing X-ray properties of very faint X-ray transients (VFXTs).

\textbf{(3) Resolving the central 10 parsec region above 10 keV (\S\ref{sec:nsc}): }  The nuclear star cluster (NSC), spanning over $r \simlt 10$ pc around Sgr A*, contains an extremely high concentration of stars and X-ray sources where the central hard X-ray emission (CHXE) was discovered by \nustar. \hexp\ will be able to resolve the CHXE in the 20-40 keV band with better spatial resolution than \nustar, and determine whether the core of the CHXE is cuspy or not. This refined hard X-ray view of the NSC/CHXE will enable testing  theoretical models on the formation of X-ray binaries  through interactions with Sgr A* BH and other stars. Moreover, \hexp\ will obtain the most pristine images of non-thermal X-ray sources in the vicinity of Sgr A* over 40 keV, including the central PWN, X-ray filaments, and unidentified hard X-ray sources.

\textbf{(4) Probing the composition of the Galactic Ridge X-ray emission (\S\ref{sec:grxe}):} The Limiting Window, 
known as one of the low extinction regions where \chandra\ resolved Fe line emission to a few hundred X-ray point sources, holds the key to unravel the composition of the Galactic Ridge X-ray emission (GRXE). 
Due to the narrow-band \chandra\ observations below 8 keV, their true plasma temperatures, possibly varying between different X-ray luminosities (which may reflect different types of CVs in the LW), remain unknown. 
\hexp\ is expected to classify a majority of the \chandra\ sources by performing X-ray spectral analysis on bright sources individually and analyzing stacked spectral data.  Consequently, we will be able to accurately measure the plasma temperatures of the X-ray sources across different X-ray luminosity ranges. Such measurements will lead to identifying which types of X-ray sources predominantly constitute the GRXE.

\textbf{(5) Surveying the central molecular zone (\S\ref{sec:molecular_clouds}):} 
Within the central molecular zone (CMZ), a handful of molecular clouds have exhibited X-ray emission. This emission consists of variable X-ray reflections of Sgr A* outbursts in the past and/or steady X-ray emission originating from cosmic-ray interactions with the clouds. In the primary observation program, \hexp\ will conduct a survey of the Sgr A complex, Sgr B1, B2 and C, aiming to map their neutral Fe fluorescent line emission and X-ray continuum emission. The initial survey data will play a pivotal role in guiding subsequent \hexp\ observations, dissecting the X-ray emission mechanisms, and gaining valuable insights into the past activities of Sgr A* and cosmic-ray distributions around the clouds.

\textbf{(6) Investigating energetic particle 
 accelerators and mapping cosmic-ray distributions (\S\ref{sec:filaments} and Renolds et al. 2023):}  
As an integral part of the major observation program aimed at exploring Galactic particle accelerators, \hexp\ will observe prominent cosmic-ray acceleration sites, including the youngest SNR in our Galaxy (G1.9+0.3), the most luminous 
 TeV sources (e.g., PWN G0.9+0.1), and star clusters (e.g., Arches) in synergy with future CTAO TeV observatory (Reynolds et al. 2023;  Mori et al. 2023). 
 Furthermore, \hexp\ will characterize the synchrotron radiation emitted by X-ray filaments in conjunction with high-resolution radio data. Through broadband X-ray spectral data of these filaments, we will create a detailed map of the distribution of TeV--PeV electrons in the GC. Ultimately, \hexp\ will explore how and where the relativistic cosmic rays are accelerated and then propagate through the GC region.

\section{\sgra \ flares} \label{sec:sgra_flare}

\sgra \ is the radiative counterpart of the supermassive BH at the center of the Milky Way. \sgra \ has been proposed as one of the energetic particle accelerators powered by accretion flow among other PeVatrons candidates in the GC, such as PWNe and star clusters \citep{HESSCollaboration2016}. However, the quiescent emission from \sgra \ remains so faint with $L_{Bol}\sim 10^{36}$  erg s$^{-1}$ and $L_X\sim 2\times10^{33}$ erg\,s$^{-1}$, that it casts doubts about its role as a persistent particle accelerator \citep{Genzel2010}. On the other hand, \sgra \ emits flares frequently, during which its X-ray luminosity can increase by up to two orders of magnitude \citep{Baganoff2001, Neilsen2013, Barriere2014, Ponti2015, Haggard2019}. Similarly, IR flares have been observed daily \citep{Genzel2003,Ghez2004}. Interestingly, whenever an X-ray flare is observed, an IR counterpart is present, whereas the opposite is not always true. These flares are considered to originate from synchrotron or synchrotron self-Compton scattering radiation emitted from accelerated electrons to high Lorentz factors ($\gamma_e\sim 10^{5\rm{-}6}$ in the synchrotron case). Typically, X-ray flares last for a few thousand seconds before they decay via radiative cooling or particle escaping. While simultaneous, multi-wavelength observations of \sgra \ flares in the IR and X-ray band are important, \hexp\ will uniquely explore the particle acceleration and emission mechanisms of the highest energy electrons produced by the supermassive BH at \sgra.  

Previous \chandra\ observations of \sgra \ flares suggest an average rate of 1-2 flares per day \citep{Neilsen2013,Nowak2012}. With fainter X-ray flares occurring more frequently, the expected number of detectable flares depends on the telescope's sensitivity. For instance, \citet{Ponti2015} detected 11 flares with {\it XMM-Newton} in 1.6 Ms until 2014, yielding a lower rate of 0.6 flares per day. Note that \hexp\ will be more sensitive to detecting fainter X-ray flares than {\it XMM-Newton}. It was reported that, since 2014, bright X-ray flares with 2--10 keV fluence above $\sim 1.7 \times 10^{-8}$ erg cm$^{-2}$ have seemed to occur more frequently, at a rate of 0.8 flares per day \citep{Mossoux2020}. 
Assuming these flare detection rates, 3.5--5 flares will be detected by \hexp\ during the 500 ks exposure allocated to \hexp\ observations of \sgra \ in the primary science program.  
We anticipate more \hexp\ observations of \sgra \ will be carried out through subsequent GO or PI-led GTO programs, jointly with other telescopes. Presently, \nustar, \xmm\ and \chandra\ observe \sgra \ each year, in conjunction with EHT and GRAVITY, with a total exposure of $\simgt 500$ ks. Assuming that similar multi-wavelength observation campaigns will be undertaken in the 2030s, it is projected that \hexp\ will detect 18--25 X-ray flares during the first five years of operation. The orbit of \hexp\ at L1  enables uninterrupted coverage of \sgra. This is a tremendous advantage for characterizing the flare lightcurve profiles, whereas previous \nustar\ observations were hampered as parts of the flares were missed due to earth occultation.  

\begin{figure}[]
\begin{center}
\includegraphics[width=0.48\textwidth]{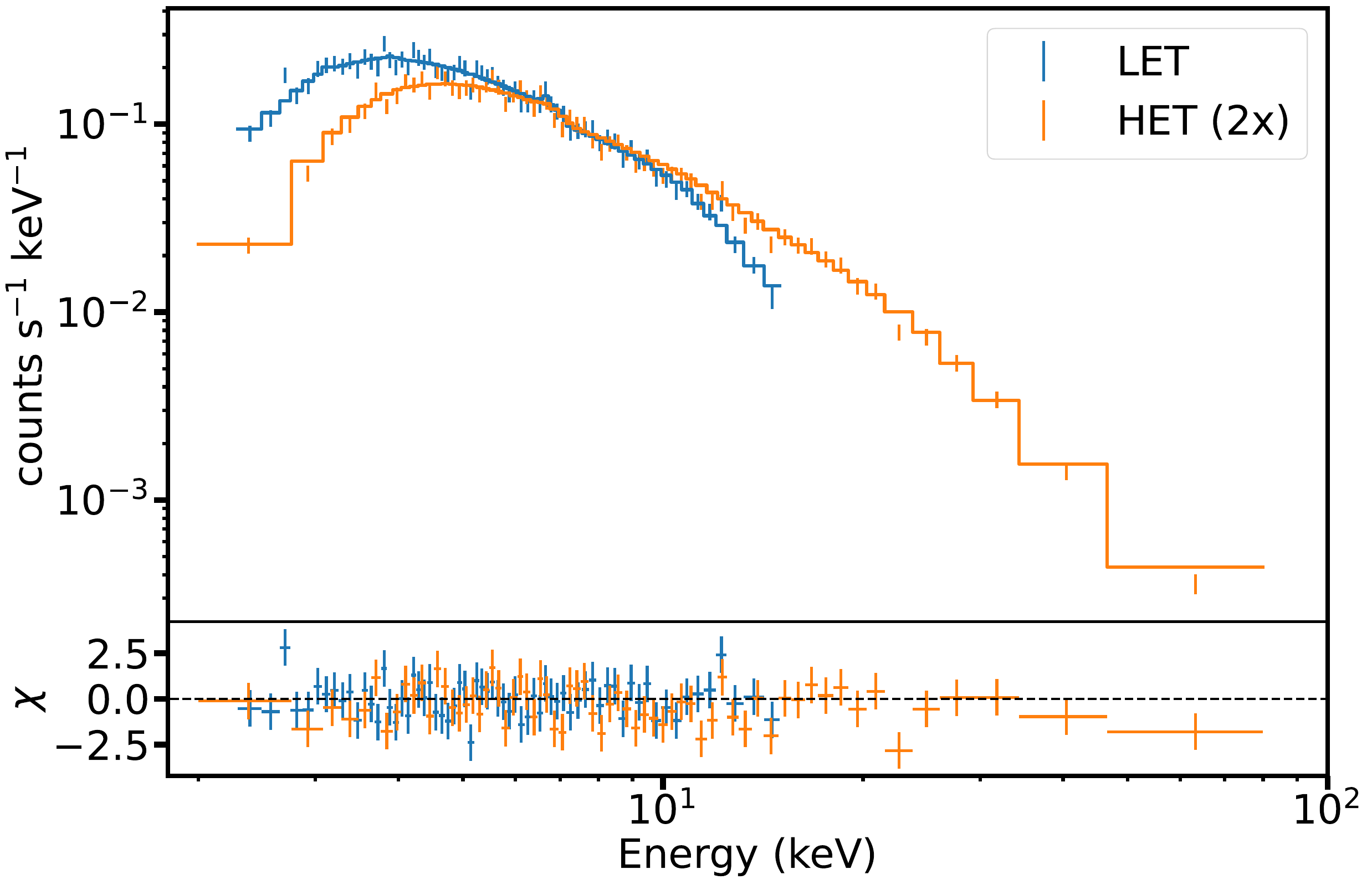}
\includegraphics[width=0.48\textwidth]{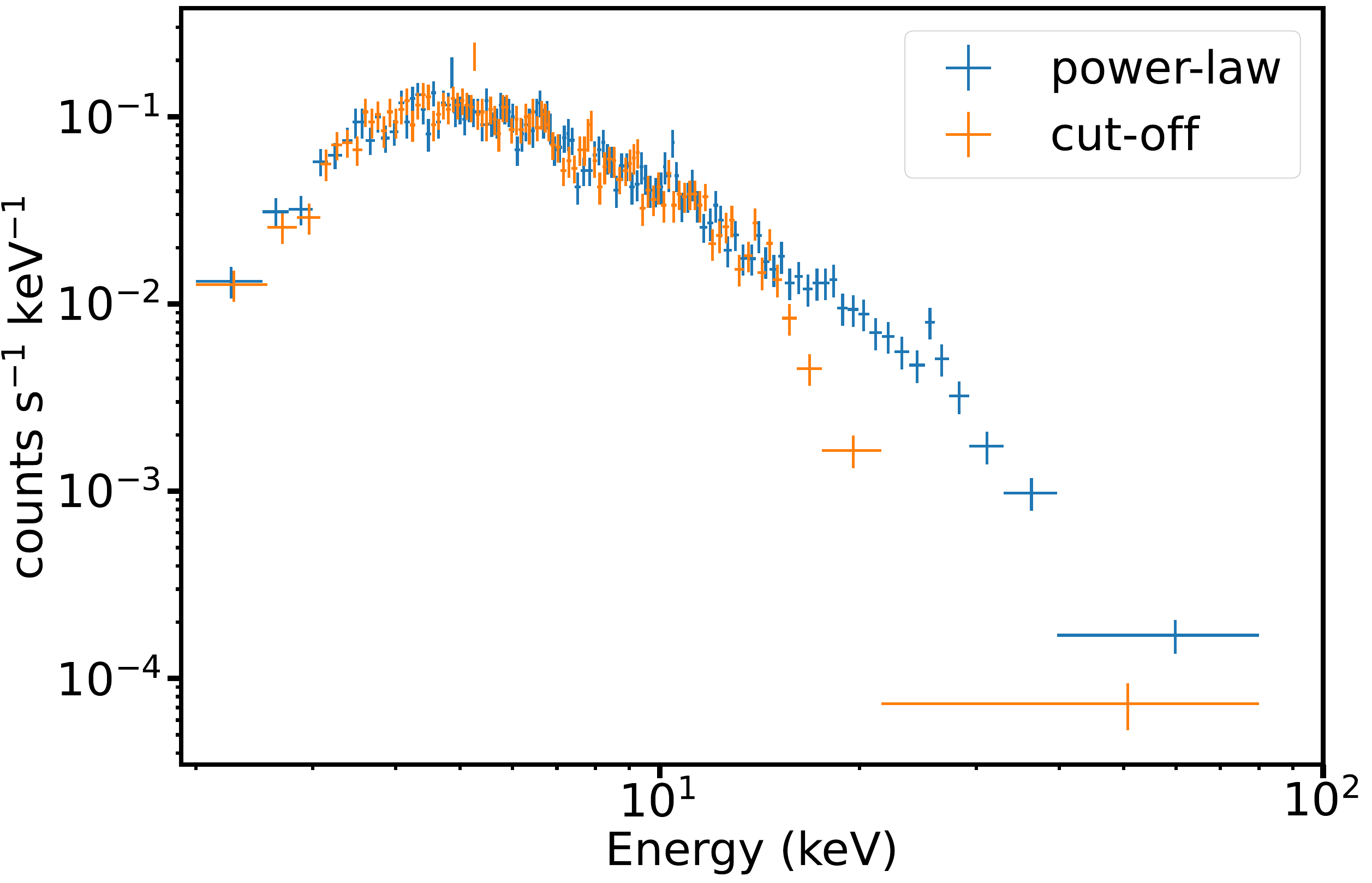}
\includegraphics[width=0.6\textwidth]{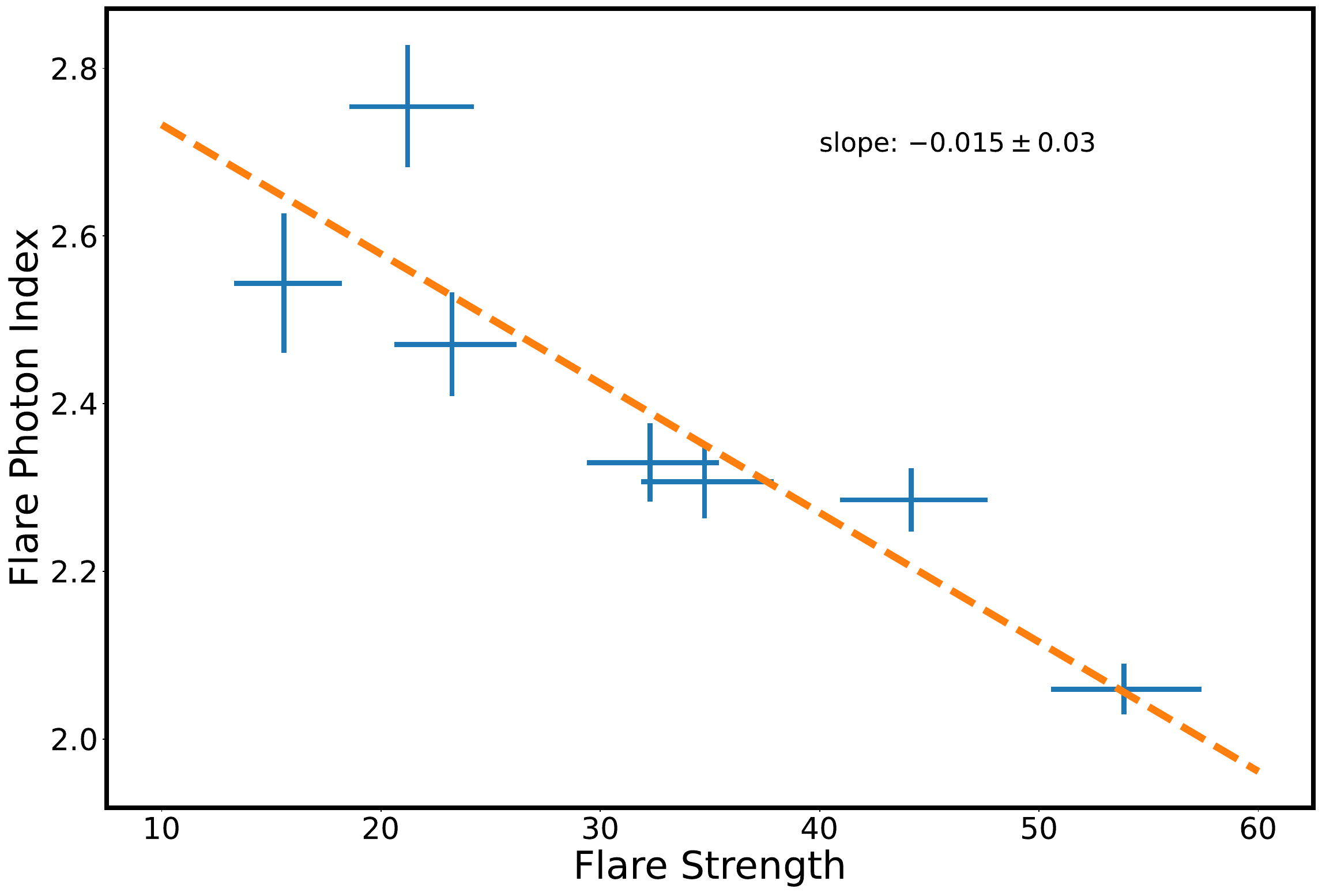}
\end{center}
\caption{{\it Top left:} Simulated \hexp\  spectra of a bright flare. The flare has a mean luminosity of 54 times the quiescent level in 2-10 keV and lasts for 5900 s.  {\it Top Right:} Simulated HET spectra of a bright \sgra \ flare, with a cutoff at 15 keV (orange) and without it (blue).  In both cases above, we adopted the astrophysical background spectra from \xmm\  observations of \sgra \ in the 2-10 keV band, and extrapolated them up to 80 keV.
{\it Bottom:} A scattered plot of the X-ray photon indices and flare strengths measured by \hexp, assuming a similar ensemble of X-ray flares as that studied in \citet{Zhang2017}. The flare strength is defined as the mean luminosity of a flare divided by the quiescent luminosity in the 2-10 keV band.} 
\label{fig:sgra_flare}
\end{figure}

The primary objective of investigating synchrotron X-ray emission during \sgra \ flares is to determine their photon indices and potential spectral cutoffs. X-ray photon indices are directly linked to the intrinsic energy distributions of electrons, while X-ray spectral cutoffs reflect the maximum energy at which electrons are accelerated or a synchrotron cooling break. These X-ray spectral characteristics are likely to vary between different flares, such as faint vs bright or short vs long flares, and \hexp\ is the most suitable for these measurements due to its broader energy coverage (compared to \chandra\ and \xmm) and higher sensitivity (compared to  \nustar).  For example, Figure \ref{fig:sgra_flare} shows simulated \hexp\ spectra of a bright X-ray flare from \sgra \ assuming different spectral shapes. 
These simulation results represent a significant advance as \hexp\ will be able to detect spectral cutoffs from Sgr A* flares, which was not achievable by \nustar\ due to the limited sensitivity. Overall, \hexp\ will provide much greater insight into the particle acceleration mechanism near the event horizon.

Furthermore, \hexp\ will eventually collect a large sample of X-ray flares from \sgra\, revealing their statistical properties. 
Based on seven bright flares observed with \nustar\ in $\simeq 1$ Ms, \citet{Zhang2017} reported a potential but unconfirmed correlation between the photon index and flux data. 
Figure \ref{fig:sgra_flare} (bottom panel) presents a scattered plot of the photon indices and the flare strengths measured by \hexp, assuming the same flux distribution observed with \nustar\ and that the potential correlation reported in \citet{Zhang2017} exists. With about five-seven bright flares, \hexp\ will be able to establish the correlation at $\gtrsim 3 \sigma$, and more significantly from a larger number of flares detectable by \hexp\ in the first few years.

\subsection{Synergy with radio, IR, and TeV gamma-ray telescopes} 
 
Simultaneous X-ray and IR observations have proven to provide a more in-depth investigation of the particle acceleration and cooling mechanisms of \sgra, as recently realized by a few such successful observations. For instance, EHT directly images the SMBH and its environment in the sub-millimeter band, potentially capturing hot spots swirling around the event horizon following X-ray flares \citep{Wielgus2022}. In the IR band, GRAVITY has pinpointed  \sgra \ and detected flares with its exceptional sensitivity of the adaptive optics. 
Since the X-ray and IR emission represent different components of the entire synchrotron radiation emitted from \sgra\ flares, multi-wavelength SED data will enable not only distinguishing between various emission models but also determining the magnetic field strengths and maximum electron energies \citep{Dodds-Eden2009, Ponti2017}. 

Previous IR and sub-millimeter observations have revealed that flares originate from an area that extends over one gravitational radius, and they orbit around $\sim5$ Schwarzschild radii from the SMBH at a non-Keplerian speed \citep{Gravity2018, Gravity2020}. However, these observations are less sensitive in determining the acceleration mechanism since the electron cooling times in these bands are orders of magnitude longer than those for X-ray-emitting electrons. Therefore, only X-ray observations can be used to identify the central engine of accelerating particles to the highest energies and whether it is powered by magnetic reconnection, shock acceleration, etc. 
As mentioned earlier, by studying bright flares through \hexp\ observations, we can directly measure the energy distribution and maximum energy of the accelerated electrons (combined with the magnetic field strength determined by IR and X-ray SED data). Tracking the evolution of the electron spectra during a very bright flare may be feasible as \hexp\ will yield ample photon statistics of X-ray flares with continuous coverage. Sensitive X-ray observations with \hexp\ can also reveal the formation and evolution of the acceleration source on a timescale comparable with the light crossing time, which is of the order of the Schwarzschild radius. These are unique opportunities that can be pursued by \hexp\ but are not feasible with the limited bandpass of \chandra\ and \xmm.

Recent \nustar\ observations suggested somehow softer photon indices of Sgr A* flares compared with those measured by \xmm\ and \chandra. This could indicate the presence of a high energy cut-off above 10 keV \citep{Gravity2021} or an artifact caused by the high background level in \nustar\ data. In contrast, \hexp\ observations of Sgr A* flares over a broader X-ray band and with significantly reduced background contamination provide  
more direct insights into how particles are accelerated at a few Schwarzschild radii from the SMBH and will help to constrain the acceleration models and their physical parameters.  If Sgr A* is indeed accelerating electrons to high Lorentz factors above $\gamma_e \sim 10^6$, which will be evident from detecting hard X-rays with \hexp, these electrons will also emit TeV gamma-rays through the inverse Compton scattering process. Thus, hard X-ray detections by \hexp\ will motivate simultaneous TeV observations with the upcoming CTAO mission and ultimately determine whether the supermassive BH at Sgr A* is a PeVatron accelerator at the present time.

\section{X-ray source population in the GC and Bulge \label{sec:gc_pointsources}}

\begin{figure}[h!]
\begin{center}
\includegraphics[width=\textwidth]{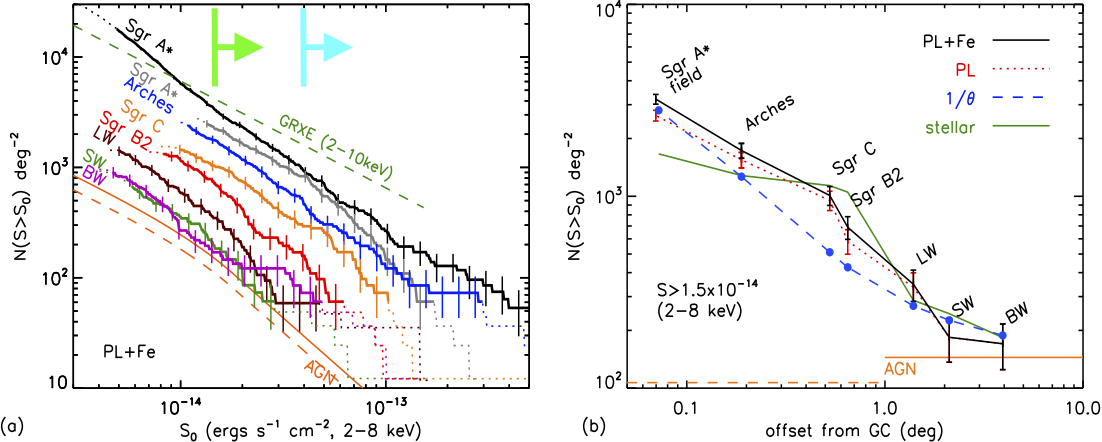}
\end{center}
\caption{logN-logS (left) and radial (right) distributions of \chandra\ X-ray sources in the GC. The plots are excerpted and modified from  \citet{Hong2016}. The 2-10 keV flux limits are indicated by the vertical cyan and green lines with arrows for source identification by individual X-ray spectral analysis (1) and classification by hardness ratio analysis described (2), respectively. Note that we assumed 100 ks exposure per source, and these conditions for source identification and classification do not apply to crowded fields such as the Arches cluster and the central 10 pc region where the source density is much higher (see the right panel).  These X-ray flux limits correspond to $L_X = 3\times10^{32}$ and $1\times10^{32}$ erg\,s$^{-1}$ at the GC distance (8 kpc), respectively. }\label{fig:ps_distribution}
\end{figure}

Extensive \chandra\ surveys have detected over 10,000 X-ray sources in the 2$^\circ \times 0.8^\circ$ GC field \citep{Wang2002, Muno2009, Zhu2018}. Even excluding foreground sources, which constitute about 25\% of the total, these \chandra\ sources represent the densest  
X-ray source population in our Galaxy.  This is illustrated in Figure \ref{fig:ps_distribution}, which shows the \chandra\ X-ray flux distributions in different regions of the GC and Bulge.  
In the GC and bulge regions, where optical, UV, and soft X-ray ($E\simlt2$ keV) observations are hindered due to significant extinction and dust scattering, the hard X-ray band provides a singular opportunity to explore the populations of compact objects and their binaries.   
Thus far, multi-decade X-ray observations of the GC, Bulge, and Ridge have revealed that (1) the X-ray point sources are predominantly mCVs, with a modest fraction of XRBs \citep{Xu2019b}; (2) a small population of non-thermal X-ray sources exist, and they are likely LMXBs or runaway pulsars \citep{Hailey2018}; (3) $\sim20$ X-ray transients have been detected, a majority of them in the central few pc \citep{Mori2021}; (4) diffuse hard X-ray emission is largely composed of unresolved CVs \citep{Hailey2016}. 
As described briefly below, these results indicate distinct spatial distributions and compositions of X-ray sources in the GC and Bulge. Identifying these X-ray sources and the underlying X-ray populations is important for testing the fundamental theories and recent N-body simulation results of XRB/CV formation in the vicinity of Sgr A*, the NSC and beyond \citep{Szolgyen2018, Generozov2018, Panamarev2019}, with implications ranging from the rates of gravitational wave events to the nature of dark matter.

(1) \underline{Magnetic CV population in the GC and Bulge:} Stacked X-ray spectra of the \chandra\ sources were well fit with an optically thin thermal plasma model with $kT \simgt 8$~keV, indicating that they are predominantly mCVs \citep{Hong2016}. 
MCVs possess WD magnetic fields strong enough to truncate or quench accretion disks, and they are generally classified into two classes: intermediate polars (IPs) and polars.   
An IP is a type of mCV that has non-synchronized orbits and WD magnetic fields ($B\sim0.1\rm{-}10$~MG) strong enough to truncate the inner accretion disk; they are copious emitters of hard X-rays ($kT \sim 20\rm{-}40$ keV). On the other hand, polars are magnetically synchronized mCVs with higher B-field strength 
(typically $B\sim10\rm{-}240$~MG) and thus possess softer X-ray spectra ($kT \sim 10\rm{-}20$ keV) than IPs due to faster cyclotron cooling.  
By contrast, non-magnetic CVs (nmCVs) have lower B-fields ($B\simlt 0.1$~MG), and their accretion disks extend to the WD surface. NmCVs exhibit softer and fainter X-ray emission ($kT \simlt 10$ keV) than those of mCVs \citep{Byckling2010}. Since only a few \% of the \chandra\ sources in the GC have been detected by \nustar\ above 10 keV \citep{Hong2016}, their source types remain elusive. To distinguish between nmCVs, IPs, and polars, it is important to characterize both broad-band X-ray continuum and Fe emission lines at 6--7 keV \citep{Mukai2017, Xu2019}.   

(2) \underline{X-ray binaries and pulsars:} A fraction of the \chandra\ sources are characterized by non-thermal X-ray spectra with no or weak Fe lines. These non-thermal X-ray sources are likely quiescent LMXBs (containing NSs or BHs) or pulsars \citep{Mori2021}.  
Subsequently, multi-decade X-ray observations of the Sgr A* region with \chandra, \xmm, \swift\ and \nustar\ revealed a cusp of LMXBs in the central pc \citep{Hailey2018, Mori2021}. Finding a concentration of pulsars in the Galactic Bulge will have significant implications for the  
origin of the GeV gamma-ray excess emission detected by \textit{Fermi}-LAT \citep{Dexter2014}.

(3) \underline{X-ray transients:} Daily \swift/XRT monitoring since 2006, combined with other frequent X-ray observations of the GC, resulted in the detection of 20 X-ray transients in the central $\sim50$~pc \citep{Muno2005, Degenaar2015}. Six of the GC transients were identified as NS-LMXBs through the detection of type I X-ray bursts or pulsations \citep{Degenaar2012}, while \nustar\ observations identified two recent \swift\ transients as outbursting BH-LMXBs \citep{Mori2019}. 
Otherwise, $\sim10$ VFXTs have been detected with peak luminosities of $L_X \sim 10^{34}$--$10^{36}$~erg\,s$^{-1}$\citep{Degenaar2015} but their identity remains elusive despite extensive  investigations by \swift-XRT \citep{Degenaar2015}.  

(4) \underline{Diffuse hard X-ray emission:} In the Bulge, Ridge and central 10 pc regions, diffuse hard X-ray emission with different plasma temperatures ranging from $kT \sim 8$ to 35 keV have been detected by \suzaku, \integral\ and \nustar\ \citep{Yuasa2012, Turler2010, Perez2019, Perez2015}. This diffuse X-ray emission is believed to originate from unresolved populations of X-ray point sources. Only a few low-extinction regions in the bulge and ridge, such as the Limiting Window (LW), have been observed by \chandra\ and resolved into hundreds of X-ray point sources, most of which remain unidentified due to the lack of hard X-ray observations. 

As we present below, \hexp\ will greatly enhance our understanding of X-ray source populations in the GC and Bulge, covering a wide range of X-ray luminosity ($L_X$). \hexp\ will conduct deep observations of prominent regions such as the Sgr A complex, the Arches cluster, Sgr B2, Sgr C, and the LW. These investigations will fully utilize the broadband spectral and timing capabilities of \hexp\ to reveal the X-ray source populations within those regions. 
Our primary focus will be on 
classifying X-ray source populations in the range of $L_X \sim 10^{32}\rm{-}10^{33}$ erg\,s$^{-1}$ at the GC distance (8 kpc), and examining how they vary with $L_X$ and distance from the GC. This unexplored parameter space of X-ray sources holds considerable interest, as we expect to unveil a number of mCVs, XRBs, and pulsars that have not yet been discovered in the GC. In contrast, a significant portion of fainter X-ray sources is likely background AGNs and coronally active dwarfs \citep{Ebisawa2005, Dewitt2010, Morihana2022}, which are unrelated to the compact object populations in the GC and Bulge.

\subsection{X-ray source population \label{sec:source_population}} 

The \nustar\ GC survey covered $\sim 2/3$ of the central $2^\circ\times0.8^\circ$ region, resulting in detecting a total of 77 hard X-ray sources above 10 keV \citep{Hong2016}. These sources  exhibit either thermal X-ray spectra with $kT > 20$ keV or non-thermal X-ray spectra with $\Gamma = 1.5\rm{-}2$, indicating that they are mCVs or XRBs, respectively \citep{Hong2016}.  
The \nustar\ data analysis of the GC sources has been largely limited  below $\sim20$ keV due to high stray-light background levels \citep{Mori2015}. 
Consequently, despite extensive studies combining \nustar\ observations with archived \chandra\ and \xmm\ data, the majority of  {\it individual} \nustar\ sources remain unidentified because their existing X-ray spectra do not allow us to distinguish between thermal (mCVs) or non-thermal sources (XRBs), and lack accurate measurements of plasma temperatures or photon indices. Furthermore, $\sim1/3$ of the $2^\circ\times0.8^\circ$ GC region, as well as the LW,  were not surveyed by \nustar\ due to substantial background contamination. In addition to identifying the known hard X-ray sources detected by \nustar, \hexp\ will survey the unexplored regions, allowing us to detect new hard X-ray sources and determine their source types.

The goals and capabilities of \hexp\ in classifying and identifying X-ray sources in the GC are specified in Table \ref{tab:point_sources} 
in the Appendix and in the 
numbered paragraphs below. The primary GC survey program involves surveying the central $1^\circ \times 0.7^\circ$ region with multiple tiled observations with a typical exposure time of $\sim100$ ks per field. Each tile will have a 50 ks exposure,
likely divided into multiple, shorter exposures for obtaining X-ray variability data. These tiles will overlap with each other to achieve $\sim100$ ks exposure at a given location in the survey area. Additionally, \hexp\ plans to conduct deeper observations of Sgr A*, Sgr A, and the LW (Table \ref{tab:obs}).

\begin{figure}
\begin{center}
\includegraphics[width=0.8\textwidth]{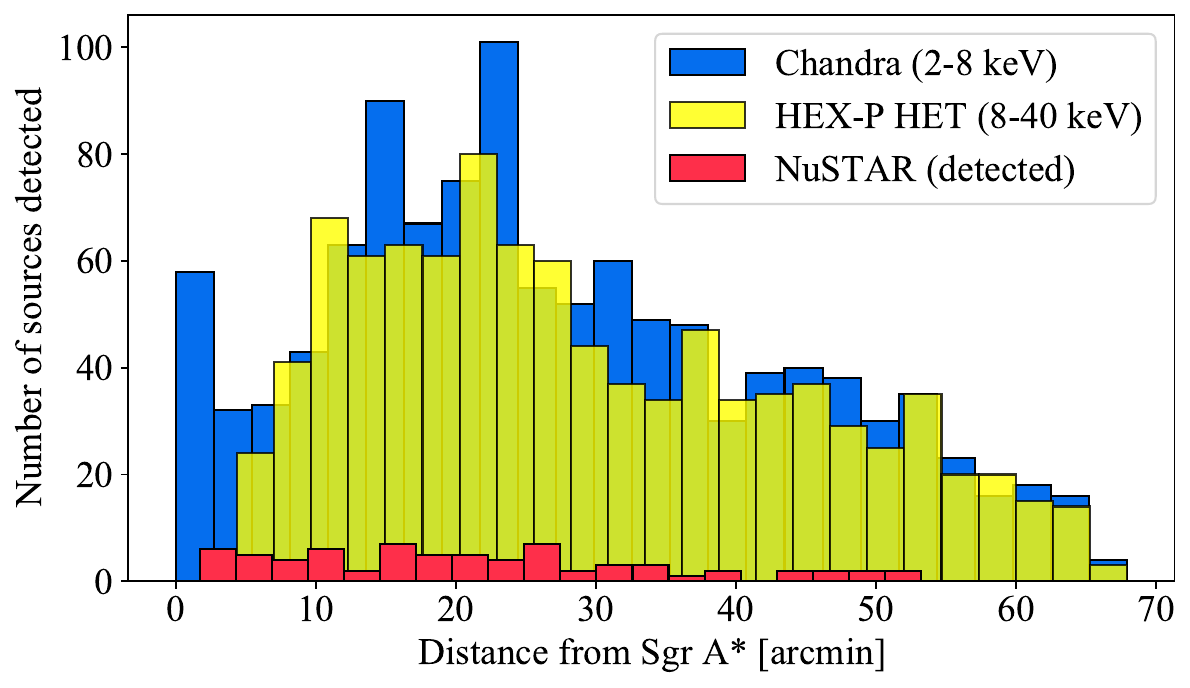}
\end{center}
\caption{The radial distributions of point sources that will be detectable by \hexp\ HET in the 8--40 keV band above the $4\sigma$ level (yellow).  The distribution only includes non-foreground sources for which the ``nearest neighbor" is at least 9.5\asec away, corresponding to half of HET's HPD \citep{Pf2023}.  For comparison, the blue histogram shows the point sources detected by \chandra\ with $> 4\sigma$ significance.   
By contrast, \nustar\ was only able to detect 70 sources (red) within the same region. }\label{fig:hexpvnustar}
\end{figure}

We conducted extensive simulations based on existing data to demonstrate the remarkable capabilities of our proposed surveys.  In order to determine the detectability of X-ray sources with \hexp, we 
adopted the X-ray flux data of 
6760 \chandra\ sources from \citet{Muno2009}, excluding foreground sources. 
We assumed that these sources have $kT = 15\rm{-}30$ keV, which are typical plasma temperatures for mCVs. Assuming a 100 ks exposure per field and $kT = 15$ keV, HET will be able to detect $>600$ sources at $>4\sigma$ significance; with a temperature of $kT = 30$ keV, that number rises to $>950$ sources within the $2^\circ \times 0.8^\circ$ GC field (Figure \ref{fig:hexpvnustar}).  HET will be able to extract X-ray photometry 
data cleanly from these sources since they are separated from each other by at least $9.5''$ (half the HET HPD),  allowing us to obtain broadband logN-logS distribution, hardness ratio, and variability data in the 2--40 keV band. This is a significant improvement compared to the 77 hard X-ray sources detected by \nustar. As described below, \hexp\ will be able to identify and classify $\sim1,000$ X-ray sources in the GC using broadband X-ray spectral and timing data, combined with the archived \chandra\ and IR data. Table \ref{tab:source_id} lists various types of X-ray sources that can be identified by extensive \hexp\ observations in the GC and other regions in our Galaxy.

\renewcommand{\arraystretch}{1.2}
\setlength{\tabcolsep}{7.pt}
\begin{table*}[h!]
{\small
\begin{center}
\caption{X-ray source classification \label{tab:source_id}}
\begin{threeparttable}
\begin{tabular}{lccccc}
\hline\hline
Source type & $L_X$\tnote{a} & Fe lines\tnote{b} & X-ray continuum & Variability & IR detection\tnote{c}
\\ 
 & [erg\,s$^{-1}$] & & & & \\  [0.25ex]
\hline\hline
LMXB  & $10^{31}\rm{-}10^{33}$ & Weak & Thermal + Power-law\tnote{e} & Yes & Yes (MS donor) \\
HMXB &  $\simgt10^{33}$ & Strong & Thermal + Power-law & Yes & Yes (O/B donor) \\ 
UCXB &  $\simlt 10^{33}$ & Weak & Thermal ($kT_{\rm BB}\sim0.1\rm{-}2$ keV) & Yes &  No (H-poor donor) \\ 
Pulsar & $10^{32}\rm{-}10^{34}$ & No & Power-law ($\Gamma \sim1.3\rm{-}1.5)$ & No & No \\ 
IP &  $10^{32}\rm{-}10^{34}$ & Strong &  Thermal ($kT\sim15\rm{-}40$ keV) &  Low & Yes (MS donor) \\
Polar & $10^{31}\rm{-}10^{33}$ & Strong  & Thermal ($kT\sim5\rm{-}15$ keV) & High & Yes (MS donor) \\ 
Non-magnetic CV & $\simlt 10^{31}$ & Strong  & Thermal ($kT\simlt 8$ keV) & Low & Yes (MS donor) \\ 
AM CVn &   $\simlt 10^{33}$ & Weak &  Thermal ($kT\simlt 1$ keV) & Yes & No (H-poor donor) \\ 
BH/NS transient & $10^{36}\rm{-}10^{38}$ & Strong & Thermal + Power-law & Yes & Yes (MS donor) \\ 
VFXT\tnote{d} & $10^{34}\rm{-}10^{36}$ & Weak & Thermal + Power-law & Yes & Yes (if MS donor) \\
\hline
\end{tabular}
\begin{tablenotes}
    \item [a] Unless otherwise noted, these values refer to typical quiescent luminosities. \item [b] Strong and weak Fe lines refer to the combined equivalent width of neutral, He-like and H-like lines at 6.4, 6.7 and 7.0 keV above $\sim 300$ eV and below $\sim100$ eV, respectively. 
    \item[c] This refers to the detection of IR counterparts \textit{at Galactic center distances}.
    \item[d] VFXTs (very faint X-ray transients) could be outbursting UCXBs. 
    \item[e] Some LMXBs, particularly low-accretion neutron stars, exhibit primarily thermal emission; others, especially quiescent BH-LMXBs, are predominantly non-thermal.
\end{tablenotes}
\end{threeparttable}
\end{center}
}
\end{table*}

\textbf{ (1) Identification of Chandra and IR counterparts: } 18 out of the 77 \nustar\ sources do not have clear associations with \chandra\ sources because the limited angular resolution of \nustar\ ($18''$ FWHM) hinders  their correlation with \chandra\ positions, as well as their IR counterparts. 
\hexp’s angular resolution, superior to \nustar’s, will allow us to identify \chandra\ counterparts to hard X-ray sources with high confidence. Consequently, \chandra’s sub-arcsecond angular resolution will enable us to find the NIR counterparts to those X-ray sources and determine their nature using the combined X-ray and NIR spectral and temporal properties of those systems (Table 2). Accreting binaries are expected to display ellipsoidal modulations in their NIR light curves, which result from the distortion of Roche-lobe-filling donor stars due to the tidal forces exerted by the compact object. 
Currently ongoing -- and upcoming -- JWST surveys will be able to detect such periodic variability even from low-mass companions, thanks to their unprecedented depth \citep{Schoedel2023}. Combining deep NIR and X-ray data will open a new window into the nature and evolution of accreting compact objects. Furthermore, unambiguous \chandra\ counterpart identification 
will enable 
joint X-ray spectral and variability analysis with archived \chandra\ and \xmm\ data, allowing us to understand the temporal properties of those sources over a timescale of decades.

\begin{figure}[h!]
\begin{center}
\includegraphics[width=1.05\textwidth]{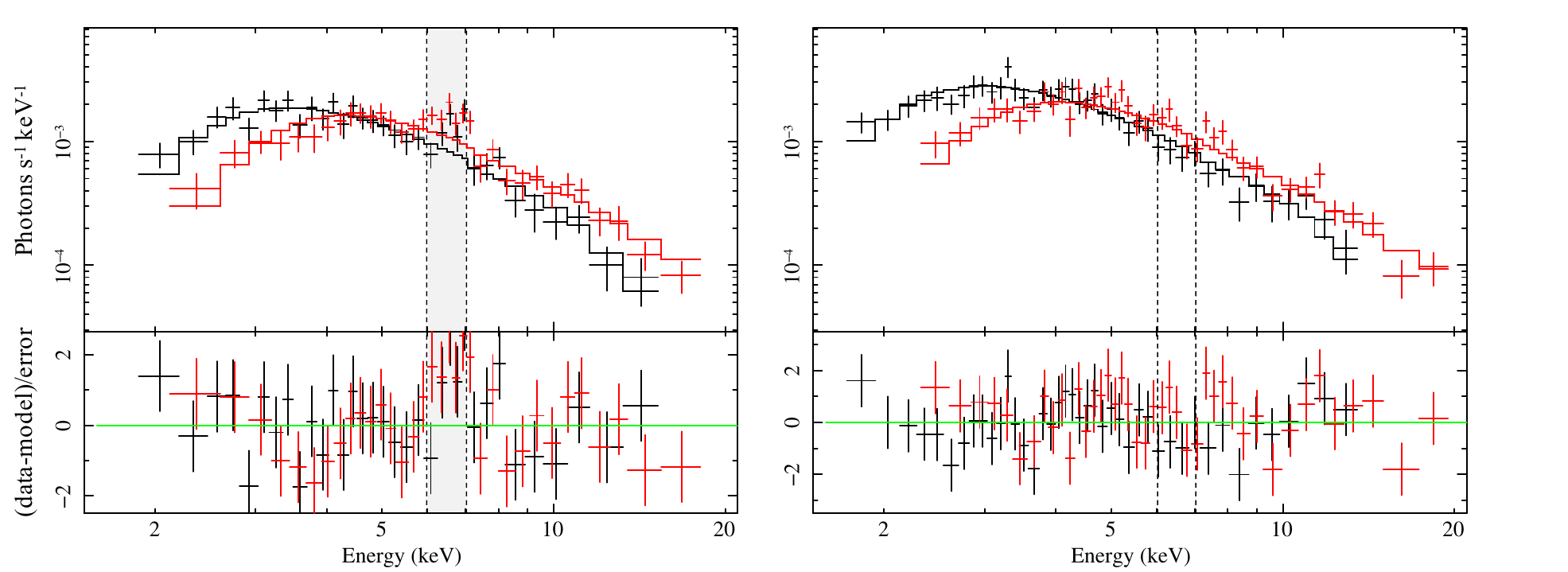}
\end{center}
\caption{Simulated \hexp\ LET (black) and HET (red) spectra of thermal (left) and non-thermal (right) X-ray sources in the GC, assuming 100 ks exposure. 
We fit an absorbed power-law model to
both sources.  
The thermal source exhibits Fe line
residuals at 6--7 keV (shaded area), which are detected with $5\rm{-}\sigma$ significance.  }
\label{fig:gc_source_sim}
\end{figure}

\textbf{(2) X-ray spectral analysis: }   
Hard X-ray detections above 10 keV suggest the \nustar\ sources could be either mCVs or XRBs \citep{Hong2016}. To accurately identify these sources, Fe line diagnostics are crucial. Typically, mCVs exhibit thermal X-ray spectra with strong Fe emission lines at 6--7 keV \citep{Xu2016}, while quiescent XRBs show non-thermal X-ray spectra with weak or no Fe lines. However, the majority of the \nustar\ sources  lack clear Fe line 
determinations 
because their \nustar\ spectra are largely contaminated by diffuse X-ray emission, which also exhibits strong Fe lines. 
With both the LET and HET  
covering the 6--7 keV band, \hexp\ 
will greatly improve Fe line diagnostics thanks to its superior angular resolution and significantly reduced background levels compared to \nustar. Figure \ref{fig:gc_source_sim} (top) illustrates how \hexp\ can discern between thermal and non-thermal X-ray models by resolving the Fe line complex region at 6--7 keV. 
For faint X-ray sources that lack sufficient source counts for individual spectral analysis, \hexp\ will be highly effective in classifying source types via hardness ratios and stacked X-ray spectra. Given its broad energy band, \hexp\ is less susceptible to X-ray absorption below a few keV, which can cause parameter degeneracy with intrinsic spectral parameters like $kT$ and $\Gamma$. Unlike the limited bandpass of \chandra\ and \xmm\ data, \hexp\ can investigate the broad-band hardness or quantile ratios, which are more  intrinsic to source types and properties. For instance, one can define hardness ratios between 3--10 and 10--40 keV bands, which are nearly unaffected by X-ray absorption and dust scattering effects.

\textbf{(3) Detection of X-ray variability and periodic signals: } 
Multi-epoch \hexp\ observations of the GC regions can provide long-term X-ray variability data over the course of days and weeks. Year-scale X-ray variabilities can be investigated by combining \hexp\ with archived X-ray data obtained by \nustar, \chandra\, and \xmm over the past decades. These long-term X-ray variability data can be useful in identifying mCV types. For example: compared to IPs, polars often display significant flux variations (by more than an order of magnitude) in the optical and X-ray band due to higher variability in mass accretion rates. In the GC and bulge regions, optical observations are hindered by significant extinction and dust scattering, whereas IR emission of mCVs is usually dominated by companion stars. 
\chandra\ observations have shown promise in detecting orbital periods from a handful of mCVs in the LW \citep{Hong2012, Bao2020}, which \hexp\ could also achieve by using several sophisticated timing methods with only hundreds of source counts. Based on the simulation-based detection rate from \citet{Bao2020}, in the case of a sinusoidal light curve, the detection completeness with significance over 99\% by GL algorithm \citep{GL1992} was generally $\gtrsim$ 20\% and 50\% for source counts $\gtrsim$ 100 and 300, respectively. Moreover, at high amplitude variation ($F_{\rm max}/F_{\rm min}\gtrsim 4$), which is typical for polars, the detection completeness will be $\sim90$\%.

\subsection{X-ray transients \label{sec:transients}}

The GC region hosts a dense population of XRBs and X-ray transients, as revealed by deep X-ray surveys and long-term monitoring over the last two decades \citep{Hailey2018, Mori2021, Muno2005, Degenaar2010}. 
In the crowded GC region, \hexp\ is uniquely suited to make these requisite hard X-ray ($>10$~keV) observations due to its superior angular resolution. \hexp\ will be sensitive in obtaining broad-band X-ray spectral and timing data of VFXTs, which belong to a unique class of X-ray transients with peak luminosities of $L_X \sim 10^{34}$--$10^{36}$~erg\,s$^{-1}$\citep{Degenaar2015}. 
Given its combination of broad energy band and fast readout time, \hexp\ is ideal for characterizing bright X-ray transients through spectral and timing analysis without significant deadtime effects.  
Some distinct features often observed in BH and NS transients, such as thermal disk emission, Comptonization, and relativistic Fe lines, will be fully characterized in \hexp\ broad-band X-ray spectral data. Fitting the broad-band \hexp\ spectra with relativistic X-ray reflection models will yield the spin measurements of a BH transient within 100~pc of the GC as demonstrated by \nustar\ observations of two BH transients in 2016 \citep{Mori2019}. Given the high concentration of XRBs in the GC, a combination of frequent X-ray monitoring (e.g.,  \textit{Star-X} or \swift-XRT) and \hexp\ ToO observations will provide an efficient way to measure BH spins from new stellar-mass BH transients (Connors et al. submitted to FrASS). In addition, \hexp\ has superb capabilities of detecting pulsations and QPO signals in a wide frequency range from mHz to kHz, thanks to the drastic reduction of background noise which results in an increase of fractional r.m.s. variability  (e.g. Bachetti et al., this volume; Alford et al., this volume).

\subsection{Pulsars \label{sec:pulsars}} 

In 2009, {\it Fermi}-LAT detected an excess of GeV gamma-ray emission extending over the central $\sim10^\circ$ of the Milky Way Galaxy \citep{Goodenough2009}. The nature of the GeV excess still remains one of the great puzzles in astrophysics, with often-conflicting studies that suggest either a dark matter origin or a large population of unresolved pulsars \citep{Hooper2011, Brandt2015, Gautam2022}. In addition, recent N-body simulations predict run-away NSs from the NSC due to their natal kicks or interactions with the supermassive BH and surrounding stars \citep{Bortolas2017}. However, despite extensive searches, no such population of pulsars or milli-second pulsars has ever been observed, resulting in the “missing pulsar problem”. 
As demonstrated by the discovery of a transient magnetar in the vicinity of Sgr A* \citep{Mori2013}, finding even a few pulsars in the GC and bulge regions will have profound implications for the underlying NS population and test the hypothesis of NS ejection from the NSC \citep{Bortolas2017, Dexter2014}. However, it is extremely challenging to detect radio pulsars in the GC due to large dispersion measures until SKA becomes fully operational \citep{Keane2015}. With the sub-millisecond timing resolution, \hexp\ is uniquely suited for finding pulsars in the X-ray band and complementary to future radio pulsar searches in the GC and Bulge. 

\begin{figure}[h!]
\begin{center}
\includegraphics[width=0.7\textwidth]{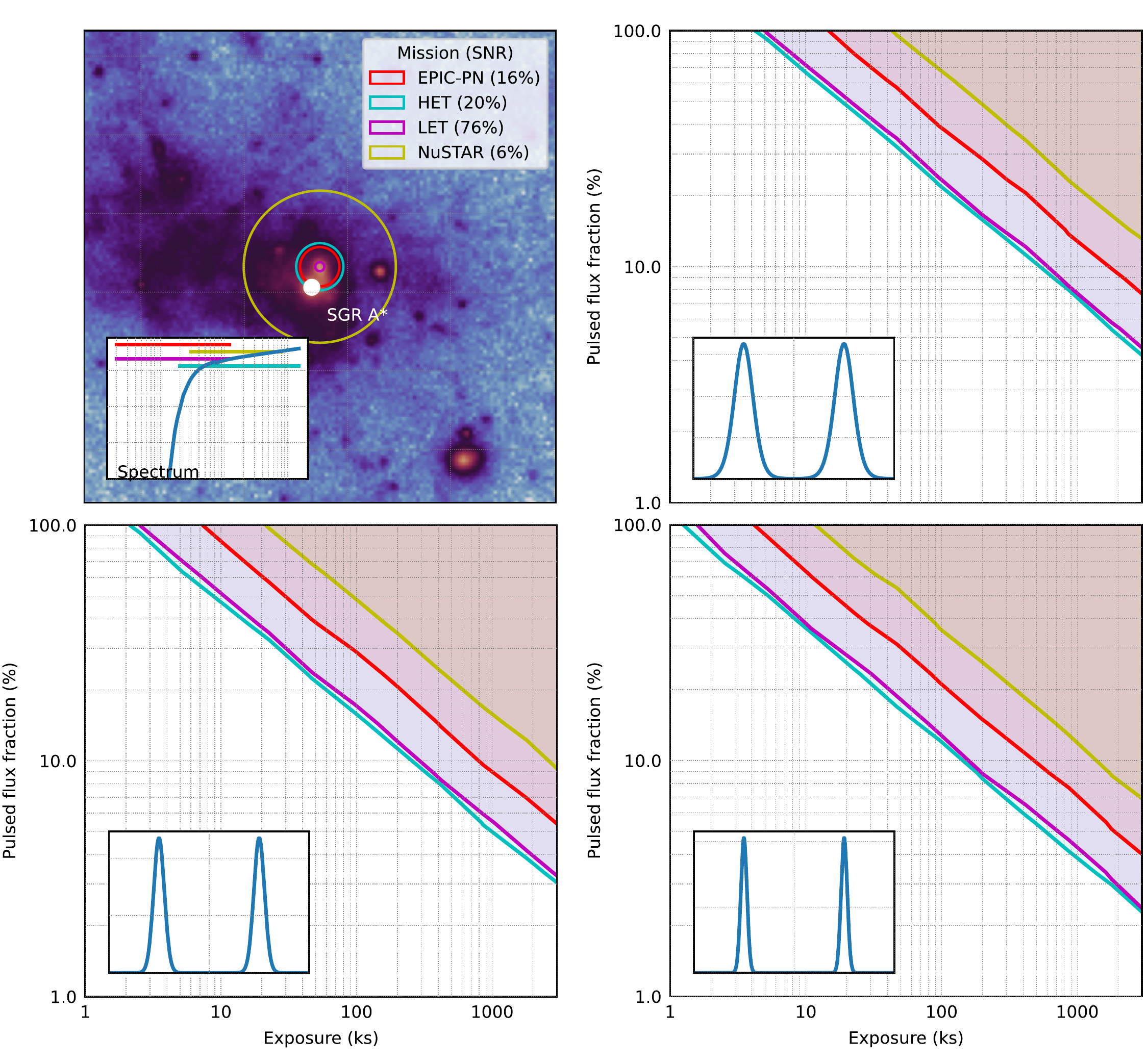}
\end{center}
\caption{
The map on the top left shows a simulated LET image of the pulsar wind nebula G359.95-0.04 and its surrounding region, using SIXTE. 
Equivalent maps were produced for the three additional instruments listed in the plot, and were used to estimate the source flux and the sky and instrumental background in the circular sources shown (calibrated on the mean HPD of the telescope).
Simulated \nustar\ and \hexp\ HET data take into account  the flux from two modules. 
The color coding is consistent in the plots (e.g., cyan for \hexp-HET).
Percentages in the legend indicate the source to background (all components) flux ratio in the region.
The shaded regions in the other plots show the pulsed flux fraction (ratio between the pulsed flux and the total flux) detectable at $>$5~$\sigma$ with different instruments, for different exposures.
}\label{fig:pulsar_simulation}
\end{figure}

There are a handful of pulsar candidates such as non-thermal, hard X-ray sources detected by \chandra\ and \nustar. For instance, a runaway NS called the ``Cannonball", detected in the radio and X-ray bands \citep{Zhao2013}, exhibited non-thermal X-ray emission extending up to 30~keV \citep{Nynka2013}. 
Furthermore, two PWNe (G359.95-0.04 and  G0.13-0.11) are associated with TeV sources \citep{Mori2015, Zhang2020}. While \chandra\ has resolved PWNe X-ray point sources within both PWNe, no pulsations have been detected yet. Since the search for radio pulsations is severely hampered by significant dispersion measures, X-ray telescopes equipped with high angular and timing resolution, such as \hexp, would be an ideal pulsar finder in the GC and Bulge. Figure \ref{fig:pulsar_simulation} illustrates the sensitivity of \hexp\ in detecting a $\sim10$\,Hz pulsation from G359.95-0.04, located $\sim10''$ away from Sgr A*. 
Higher or lower pulse frequencies would change this plot only slightly, due to a weak dependence of significance on the number of trials used in a blind search
\footnote{The number of spectral bins, and so the number of trials, is proportional to the frequency and, for the FFT, to the observation length (because $\Delta\nu=1 / T$)}
\footnote{For example, a $Z^2_2=100$ is a $8.22\,\sigma$ detection with 10,000 trials, and a $7.65\,\sigma$ detection with 1,000,000 trials.}.

Compared to \xmm\ EPIC-PN and \nustar, 
\hexp\ will be far more sensitive in detecting pulsars due to its $< 1$ msec temporal resolution, high angular resolution, and hard X-ray coverage ($>$ 10 keV), where pulsar emission is more dominant over the PWN and thermal diffuse X-ray emission. During the planned Sgr A* observations (500 ks total), both LET and HET will be able to detect a pulsation signal at $> 5\sigma$ level if the pulsed fraction is higher than $\sim10$\%. For other pulsar candidates, such as the Cannonball and G0.13-0.11, we anticipate that pulsations will be more easily detectable due to lower background rates from the surrounding regions. 
We also note that HEX-P, being in a Lagrangian point, will also guarantee much longer stretches of uninterrupted observations with little or no periods of occultation or flaring, which will improve stability for the study of slow variability and the detection of the slowest pulsars (e.g. RCW 103 with a 6.7-hr pulsation \citep{Deluca2006}). 

\subsection{Probing the distribution of compact object binaries in the nuclear star cluster \label{sec:nsc}} 

The central 10 pc region of our Galaxy contains a dense concentration of stars and compact objects around the SMBH at Sgr A*. Within the NSC, which boasts the densest stellar environment in our Galaxy, \chandra\ detected a few hundred X-ray point sources \citep{Zhu2018}. Above 20 keV, \nustar\ discovered diffuse hard X-ray emission spanning an area of approximately 8 pc $\times$ 4 pc, which coincides with the NSC \citep{Perez2015}. This emission, known as the central hard X-ray emission (CHXE), is predominantly attributed to an unresolved population of mCVs with a mean WD mass of $M_{\rm WD} \sim 0.9 M_\odot$ \citep{Hailey2016}. However, the core of the CHXE/NSC within the central few pc region could not be resolved by \nustar\ due to the limited angular resolution and presence of hard X-ray filaments.  

\begin{figure}[h!]
\begin{center}
\includegraphics[width=0.48\textwidth]{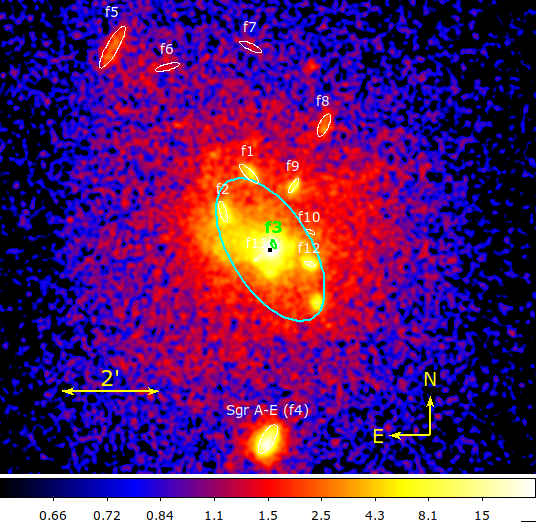}
\includegraphics[width=0.48\textwidth]{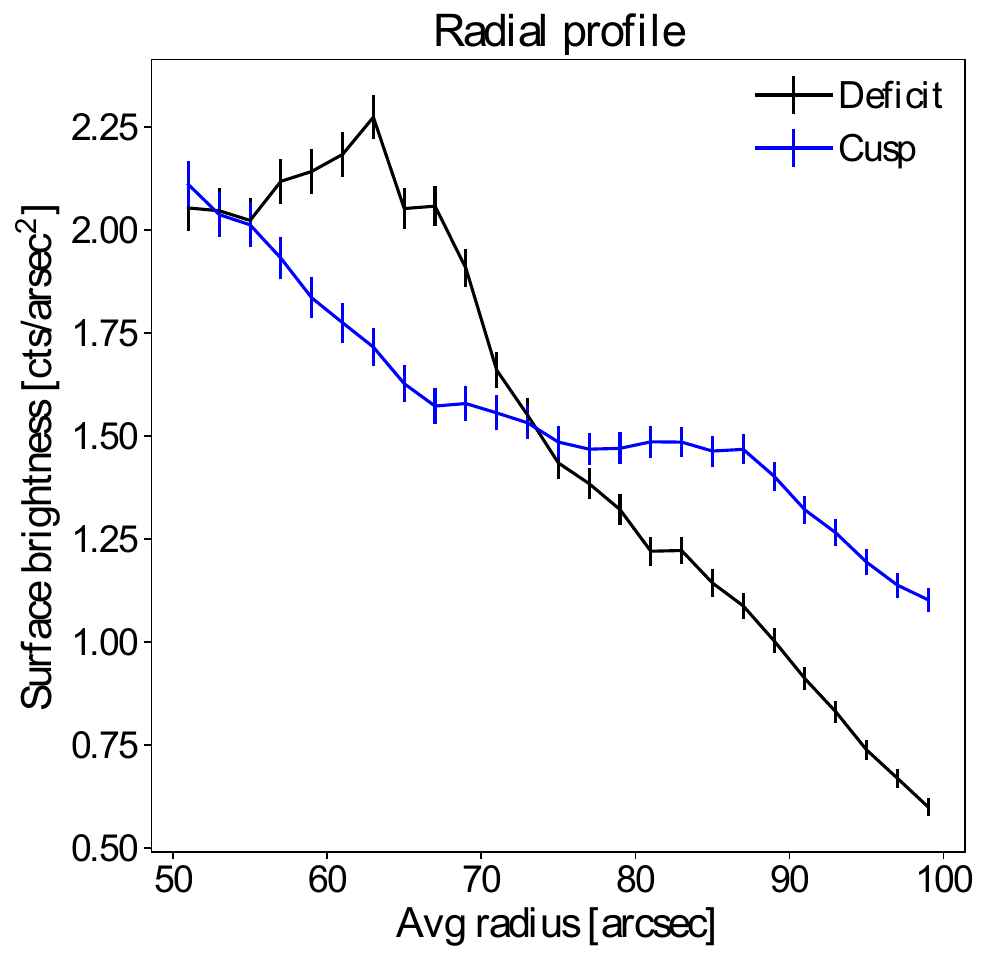}
\end{center}
\caption{Left: Simulated \hexp/HET image of the central 10 pc region in the 
20--40 keV band. Sgr A* and the CHXE are indicated by a black dot and a cyan ellipse, respectively. X-ray filaments are denoted by f1, f2, f3  etc. A complete list of the corresponding filament names and their spectral parameters can be found in Table \ref{tab:filslist} in the Appendix. Right: Simulated radial profiles of hard X-ray emission (20--40 keV) in the central 4 pc, assuming a cusp (blue) or deficit (black) of 
hard IPs at $r< 3$ pc around Sgr A*. }\label{fig:nsc_profiles}
\end{figure}

The central $\sim3$ pc region corresponds to the influence radius of Sgr A*, where the immense gravity of the SMBH affects the motion of stars and compact objects. Since NSCs in other galactic nuclei cannot be spatially resolved in the X-ray band due to Mpc distances, the NSC in the GC offers a unique opportunity to study its X-ray source compositions with \hexp. 
The gravitational interactions between Sgr A* and surrounding stars significantly boost the rates of binary formation;  
evaporation is also significantly enhanced within $r \simlt 3$ pc. As a result, this region should be largely inhabited by binaries with more massive compact objects, such as stellar-mass BHs, and with tighter orbits, as suggested by recent \chandra\ studies of the non-thermal X-ray sources detected in $r \simlt 1$ pc \citep{Hailey2018, Mori2021}. If the CHXE/NSC is indeed composed of hundreds of mCVs, an extremely intriguing question is whether its core shows a lower or higher concentration of X-ray point sources (predominantly mCVs), compared to the cuspy stellar distribution revealed by IR observations \citep{Schodel2018}. 
MCVs with heavier WD masses and shorter binary separations are more likely to survive in the NSC core. While heavier WD masses make X-ray spectra harder due to higher shock temperatures in their accretion columns, smaller binary orbits should  
correspond to reduced mass accretion rates \citep{Hillman2020}, and thus lower X-ray luminosities. A hard X-ray view of the NSC and its core within $r < 3$ pc ($< 75''$) with higher angular resolution than \nustar\ may reveal energy-dependent spatial profiles above 10 keV, where diffuse soft X-ray emission from the SNR Sgr A East becomes negligible.  
Depending on the (unknown) WD mass and binary separation distributions within $r < 3$ pc, \hexp\ may unveil a sharply lower or higher hard X-ray flux compared to the CHXE, which extends out to $r \sim8$ pc.  
In the NSC region, a unique population of mCVs, whose thermal X-ray emission is typically characterized by plasma temperatures $kT \sim 10\rm{-}40$ keV, can be probed most sensitively in the 20-40 keV band. Figure \ref{fig:nsc_profiles} shows simulated radial profiles of 20-40 keV hard X-ray emission resolved by \hexp\ for two possible scenarios representing a lower ("deficit") and higher ("cusp") concentration of mCVs within $r < 3$ pc. The two radial profiles are clearly distinct from each other, thanks to the higher angular resolution of \hexp, which can resolve the CHXE \textit{and} its core from other non-thermal X-ray sources.  

\subsection{Resolving the composition of the GRXE through the Limiting Window \label{sec:grxe}} 

While a majority of the thousands of sources detected by \chandra\ in the GC are believed to be mCVs \citep{Muno2009}, their specific source types, average WD masses, and their dependence on the X-ray luminosity ($L_{\rm X}$) and radial distance from Sgr A* are still largely unknown. It is the study of diffuse X-ray emission that has revealed the presence of distinct populations of CVs between the GC and ridge.  
In contrast to the CHXE, which has $kT \sim 35$~keV, the Galactic ridge exhibits softer diffuse X-ray emission with $kT\sim8$--15~keV \citep{Turler2010, Yuasa2012, Perez2019}. 
Several studies of this diffuse X-ray emission suggest that the softer ridge X-ray emission is due to an 
unresolved population of polars and/or nmCVs \citep{Xu2016, Hailey2016}. 
However, it is challenging to separate the diffuse X-ray emission into distinct CV populations (with varying $L_{\rm X}$ and $kT$ distributions) solely through analysis of diffuse hard X-ray continuum or Fe line emission.

The Limiting Window (LW), located 1.4$^\circ$ south of Sgr A*, is a region with low extinction that allows for the study of X-ray sources in the GC and Bulge regions without significant obscuration. A deep exposure of the LW with \chandra\ has resolved a large portion of the Fe line emission ($\gtrsim$80\%) into $\sim300$ point sources \citep{Revnivtsev2009}. This discovery has been considered a crucial indication of the point source nature of the GRXE, as opposed to being truly diffuse in origin. However, it is still unclear what types of point sources constitute the GRXE. 
\citet{Revnivtsev2009} suggest that relatively faint coronal sources, such as active binaries (ABs), may constitute a major portion of the GRXE. On the other hand, it was suggested that relatively bright accretion sources, particularly mCVs, dominate the GRXE \citep{Hong2012, Schmitt2022}. 
These conflicting findings highlight the need for further investigation to determine the contribution of different source types in the LW and shed light on the understanding of the X-ray emission in the Bulge and Ridge regions. 
Indeed, the identification of the bright X-ray sources in the LW is crucial, and their individual spectral and timing properties can provide valuable insights. 
For instance, it is difficult
to identify soft coronal sources like ABs in other typical GC and Bulge regions due to heavy obscuration, while  the  bright point source population in other low extinction windows in the Bulge such as Baade's window is dominated by AGNs (Figure~\ref{fig:ps_distribution}). 
\hexp's hard X-ray coverage above 10 keV can decisively determine the coronal vs accretion
nature of X-ray emission from these sources. 

\begin{figure}[h!]
\begin{center}
\includegraphics[width=\textwidth]{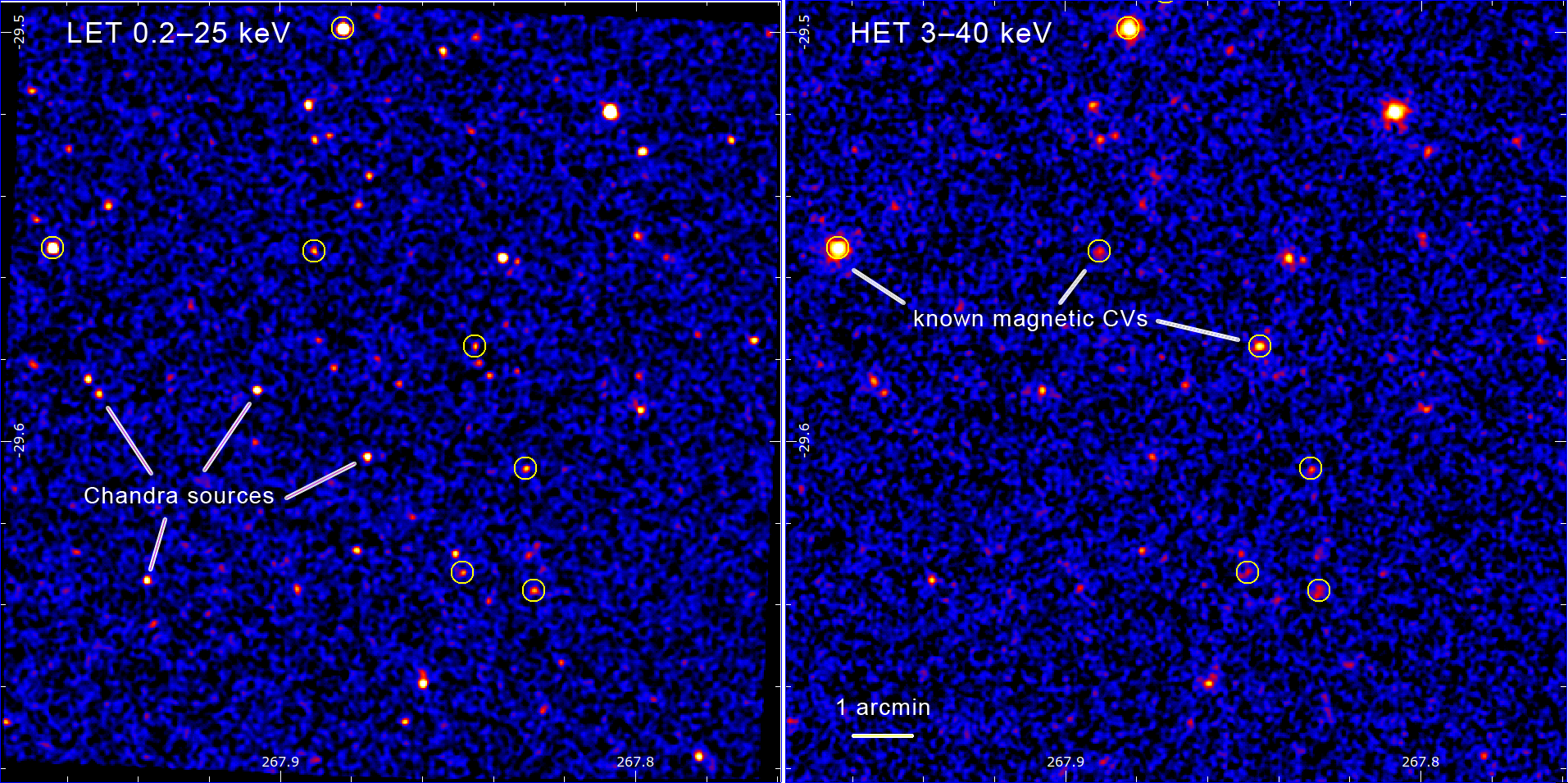}
\end{center}
\caption{Simulated \hexp\ LET (left) and HET (right) images of a section of the LW based on Chandra source positions and fluxes \citep{Hong2009} assuming an absorbed APEC model ($kT=15$ keV). Magnetic CVs identified in \citet{Hong2012} are encircled in yellow.}\label{fig:lw_images}
\end{figure}

For our simulations, we considered different X-ray source compositions in the LW. For example, we assigned $kT = 8, 15$ or 35 keV to the X-ray sources in the LW. 
Figure \ref{fig:lw_images} shows simulated LET (0.2-25 keV) and HET (3-40 keV) images of a section of the LW. Known mCVs from \citet{Hong2012} were given their best-fit model parameters specific to each source, while all other \chandra\ sources within this region were assumed to have spectra characterized by an absorbed APEC model with $kT = 15$ keV.  \hexp\ is expected to detect 53 and 29 sources above $4\sigma$ significance among 235 and 261 \chandra\ sources used as an input to SIXTE simulation for LET and HET, respectively. For different plasma temperatures ($kT =8$ and 35 keV), we found that \hexp\ can detect a similar number of \chandra\ sources in broad energy bands beyond 8 keV. 
Eventually, broad-band X-ray spectral data obtained by  \hexp\ will allow us to determine the X-ray source composition in the LW. More specifically, we aim to measure (1) plasma temperatures of the known mCVs, (2) the mean plasma temperature of other bright X-ray sources ($L_{X} \simgt 10^{32}$ erg\,s$^{-1}$), and (3) the mean plasma temperature of faint X-ray sources ($L_{X} \simlt 10^{32}$ erg\,s$^{-1}$). For group (1), we will analyze \hexp\ and archived \chandra\ spectra of the known mCVs individually. 
For groups (2) and (3), stacking their LET and HET spectra, along with archived \chandra\ data, will allow us to measure their mean plasma temperatures. In all cases, the hard X-ray coverage by \hexp\ is crucial to determine their plasma temperatures accurately (with $\sim20$\% error), which can be compared with those measured in the Bulge (8 keV), Ridge (15 keV) and CHXE (35 keV). It is possible that the three groups of X-ray sources may reveal different plasma temperatures. For instance, group (1) and (2) could have higher temperatures $kT\sim 35$ (predominantly IPs) and 15 keV (predominantly polars), respectively, while group (3) may have lower temperatures of $kT\simlt8$ keV (predominantly nmCVs and ABs) for (3). This is how  \hexp\ is capable of dissecting the composition of X-ray source populations in different luminosity ranges. Moreover, \hexp\ may be able to detect orbital periods from more X-ray sources in the LW as an alternate  way of classifying them as polars or IPs  \citep{Hong2012, Bao2020}.

\section{Diffuse X-ray sources in the Galactic Center \label{sec:diffuse_sources}} 

The GC hosts a zoo of various, unique types of diffuse X-ray sources, containing both powerful particle accelerators (e.g., Sgr A* BH, SNRs, PWNe, star clusters) and targets bombarded by relativistic cosmic rays and X-rays (e.g., molecular clouds and filaments). \hexp\ will survey all primary particle accelerators, molecular clouds (MCs), and X-ray filaments to obtain a whole picture of the CR spatial and energy distributions in the GC. Broad-band X-ray spectroscopy and excellent spatial resolution of \hexp\ are crucial for distinguishing thermal and non-thermal diffuse X-ray emission. Combined with the CTAO GC survey, \hexp\ will obtain the most accurate multi-wavelength SED data over the X-ray and TeV bands. 

\subsection{Molecular clouds \label{sec:molecular_clouds}} 

While the SMBH at Sgr A* is currently dormant, only occasionally emitting X-ray flares (see \S4), a series of X-ray observations of the CMZ that Sgr~A* used to be more active in the recent past. Following the first detection of hard X-ray emission from Sgr~B2, it was suggested that the molecular cloud (MC) reflected bright X-ray flares, presumably emitted from Sgr~A* \citep{Sunyaev1993}. The X-ray reflection exhibits Compton scattering, fluorescent Fe emission lines at 6.4 and 7.06 keV, and photo-absorption.  
In addition, both the X-ray continuum and Fe fluorescent lines have been observed to decrease over the last two decades as the X-ray flare propagated through the Sgr B2 cloud \citep{Koyama1996, Sunyaev1998, Nobukawa2011, Terrier2010, Terrier2018, Zhang2015}. Propagation of X-ray echoes is also detected from other MCs in the CMZ \citep[e.g.][]{Muno2007, Ponti2010, Clavel2013, Ryu2013, Chuard2018, Terrier2018}. High energy observations with {\it INTEGRAL} (Sgr B2) and \nustar\ (Sgr~A and B2) have shown that the reflection signal extends above 10 keV and has been detected up to 100~keV \citep{Revnivtsev2004, Terrier2010, Zhang2015, Mori2015, Kuznetsova2022}. \nustar\ played a crucial role in resolving hard X-ray continuum emission from Sgr B2 and Sgr A complex clouds, while Sgr~C has not been observed with \nustar\ due to the high background contamination \citep{Mori2015, Zhang2015, Rogers2022}. 

While the variable X-ray emission is primarily caused by single Compton scatterings, another non-variable X-ray component may emerge after the X-ray light front passes the clouds. This ``constant" X-ray emission component may be caused by (1) multiply-scattered X-ray photons confined within dense cloud cores 
 \citep{Sunyaev1998, Molaro2016, Khabibullin2020, Sazonov2020} and/or (2) non-thermal X-ray emission and collisional ionization by cosmic-rays  \citep{Tatischeff2012}. 
Recent X-ray observations indicated that clouds in Sgr~B2 \citep{Kuznetsova2022, Rogers2022} and the Arches cluster \citep{Kuznetsova2019} may have entered the constant X-ray emission phase. 
Diffuse TeV gamma-ray emission was observed to coincide with the MCs in the GC region, indicating that the clouds are bombarded by energetic CRs \citep{Aharonian2006Nat, Eldik2008, Beilicke2012, Sinha2022}. 
Overall, extensive X-ray observations over the last two decades revealed that the MCs exhibit different time evolution and X-ray properties, likely due to multiple X-ray outbursts from Sgr A* and different cosmic-ray bombardment mechanisms. 

\begin{figure}
\begin{center}
\includegraphics[width=0.7\textwidth]{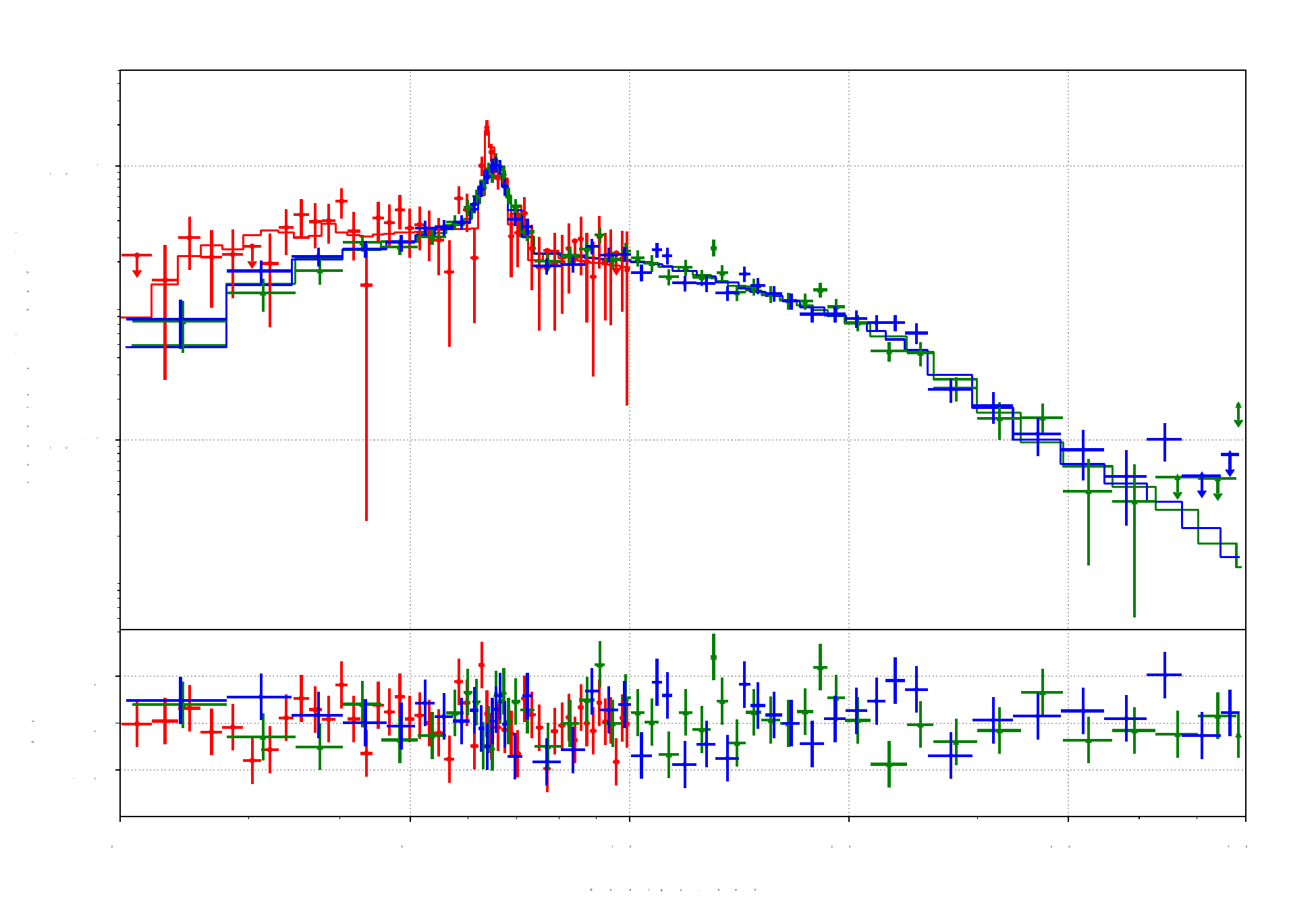}
\end{center}
\caption{Simulated \hexp\ spectra of the Sgr B2 core ($r<90''$) based on our phenomenological model \citep[see text and ][]{Zhang2015} obtained with 100~ks exposure time. The model was corrected for foreground absorption and consisted of a thermal background emission with a temperature of $kT=1.9$~keV, an intrinsically absorbed power-law with a photon index of $\Gamma=1.9$ and two Gaussian emission lines Fe~K$\alpha$ (6.4~keV) and K$\beta$ (7.06~keV), with an unabsorbed non-thermal flux in the 10--40~keV energy band of $F_{10-40~keV}=1.9\times10^{-12}$~erg~cm$^{-2}$~s$^{-1}$ \citep[other parameters are listed in table~2 in ][]{Zhang2015}. 
Red points correspond to the simulated LET data (2--10~keV), blue and green to the HET data (2--70~keV). Arrows show $2\sigma$ upper limits. The total model is represented by stepped solid lines of the colors corresponding to the data. 
}  
\label{fig:sgrb2_100ks}
\end{figure}

\hexp\ will survey all prominent MCs, including Sgr A, B1, B2 and C, in the primary GC observation program (Table~\ref{tab:obs}). The \hexp\ survey will capture the X-ray landscape of the MCs through both the Fe fluorescent line at 6.4 keV and non-thermal X-ray continuum emission in the early 2030s. \hexp\ will likely conduct follow-up observations of these clouds later to monitor X-ray variabilities. All MC complexes with X-ray variability will be covered with at least 100~ks exposure. Detection of  
changes in the morphology and spectral parameters of the MC emission will determine whether X-ray echoes are still propagating through the clouds or the cosmic-ray bombardment component is dominating. The significantly lower stray-light background will allow us to investigate the hard X-ray morphology of the CMZ with \hexp\ far more precisely than \nustar. To demonstrate the capability of \hexp\ on the MC science, we present simulation results for Sgr~B2, whose emission has been decreasing since the 1990s and has now likely reached a faint constant level. We adopted the latest \nustar\ \citep{Zhang2015,Rogers2022} and {\it INTEGRAL} \citep{Kuznetsova2022} observation results for XSPEC simulations \citep{Zhang2015}. Fig.~\ref{fig:sgrb2_100ks} shows the simulated \hexp\ spectra of the Sgr B2 core. 
Our simulation demonstrates that the non-thermal X-ray component will be detected at $\sim17\sigma$ significance, and the photon index and the equivalent width of the 6.4~keV emission line will be measured with $\leq10\%$ error.   

\begin{figure}
\begin{center}
\includegraphics[width=\textwidth]{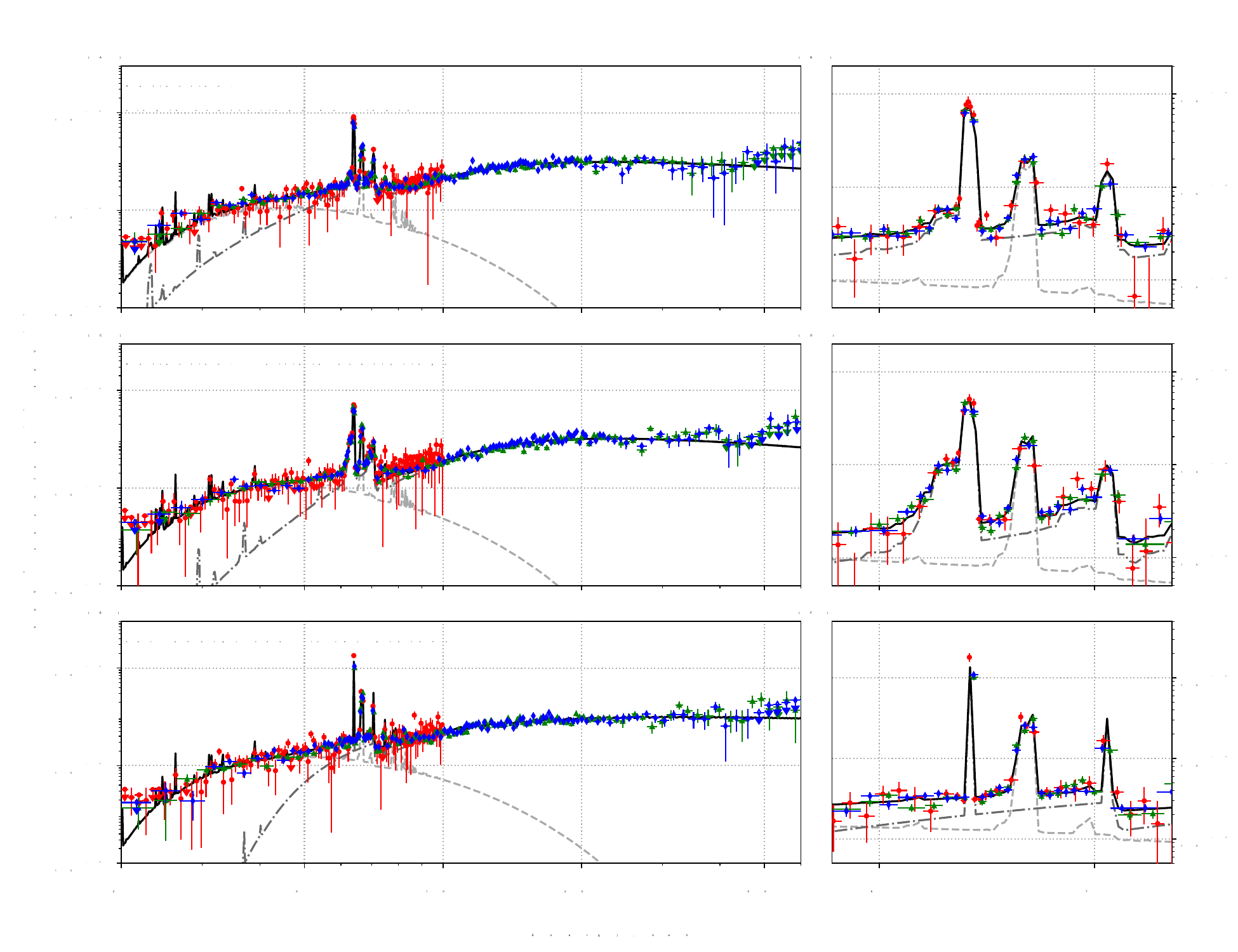}
\end{center}
\caption{Sgr~B2 spectra simulated based on the CREFL16 (panels A,B), CREFL16 multiple scatterings only (C,D), LECRp (E,F) models in the 2--70~keV energy band with 300~ks exposure. Arrows show $2\sigma$ upper limits. The right panels represent the strongest emission lines of the Sgr~B2 spectrum. Red points correspond to the simulated LET data in the 2--10~keV energy band, blue and green to the HET data in 2--70~keV. The right panels show the Fe~K$\alpha$ and K$\beta$ emission lines, at 6.40 and 7.06~keV energies, respectively, caused by non-thermal processes. Between them, there is the Fe 6.7~keV emission line related to the thermal emission component. For illustration purposes, the spectra plotted were further grouped to reach either 7$\sigma$ or 7 bins of the fitted spectra per bin.}
\label{fig:sgrb2_sim}
\end{figure}

Furthermore, we investigated whether \hexp\ can discriminate between different emission models based on broad-band X-ray spectral data. We considered two main scenarios: (1) reflection of 
past X-ray flares and (2) non-thermal bremsstrahlung emission and collisional excitation by low-energy cosmic-ray protons (LECRp). In the reflection scenario, the spectrum can be dominated by single or multiple scatterings, depending on the location of the X-ray light front.  
To clarify whether these scenarios can be discerned by \hexp, we simulated Sgr~B2 spectra for three models using 2013 \nustar\ flux measurements;  
the same low-energy component (absorption and astrophysical background) and either LECRp parameters \citep[model described in][]{Tatischeff2012} from \nustar\ \citep{Zhang2015} or reflection parameters \citep[CREFL16 model from][]{Churazov2017} from {\it INTEGRAL} observations to account for the emission of Sgr~B2 at high energies \citep{Revnivtsev2004,Kuznetsova2022}.  
Fig.~\ref{fig:sgrb2_sim} illustrates the three different models and corresponding spectra simulated for the $90''$-region of Sgr~B2. 

The Compton shoulder is clearly present below the Fe fluorescent lines (both for K$\alpha$ and K$\beta$) for the reflection models \citep{Khabibullin2020}, while the LECR spectra exhibit only the narrow emission lines \citep{Tatischeff2012}. 
The multiple-scattering case in the middle panel can be identified well by the large line ratio ($\sim1$) between the Compton shoulder and 6.4 keV line, which should be $\approx1$ as a result of multiple Compton scatterings \citep{Khabibullin2020}.

Overall, \hexp's Fe line spectroscopy and broad-band X-ray data can distinguish between the X-ray reflection and cosmic-ray models unambiguously. \hexp's spatial and spectral resolutions will allow us to trace the X-ray morphology evolution and characterize the Fe line profiles in detail, respectively. \hexp's broadband X-ray spectral data will lead to measuring the spectral index of past X-ray flares from Sgr A* and bombarding CRs. In the former case, this is a unique opportunity to investigate the mechanism of Sgr A* X-ray flares that occurred a few hundred years ago. 
For the latter, we can map out the population of CRs in the GC region in conjunction with future TeV observations by CTAO. 

\subsection{X-ray filaments \label{sec:filaments}}

A unique phenomenon in the GC region is the existence of numerous radio and X-ray filaments \citep{Yusef1984, Morris1996}. The origin and formation of these filaments have been a long-standing question for decades. 
Within the central $\sim 2^{\circ}$ region, MeerKAT radio surveys have revealed over 150 filaments, which seem to be associated with bi-polar radio bubbles \citep{Heywood2019, YZ2023}. The emission mechanism of non-thermal radio filaments has been pinned down to synchrotron emission thanks to the detection of linear polarization. Within the filaments, the magnetic field strengths are found to be 1--2 orders of magnitude higher than the surroundings at $B\sim1$~mG and aligned with the major axis of a filament (e.g. \citep{Morris1996}).
High-resolution JVLA observations further revealed complex sub-filaments entangled with each other within many of the radio filaments \citep{Morris2014}.
It suggests that radio filaments are magnetic structures, where strong and highly-organized magnetic fields trap GeV electrons and produces synchrotron emission \citep{Zhang2014}.

At higher energies, filaments at smaller spatial scales have also been detected in the X-ray band. About 30 parsec-long X-ray filaments have been detected so far, some of which have radio counterparts \citep{Muno2008, Lu2008, Johnson2009, Zhang2014, Ponti2015, Zhang2020}. 
Out of $\sim30$ known X-ray filaments, \nustar\ was able to detect only the four brightest ones \citep{Zhang2014, Mori2015, Zhang2020}. The brightest X-ray filament Sgr A-E exhibits a featureless power-law spectrum extending up to 50~keV, which requires electron energies of multi-hundred TeV electrons that are most likely produced by hadronic interactions between primary PeV protons and nearby MCs  \citep{Zhang2020}. Therefore, mapping out even a subset of the X-ray filaments could serve as a powerful way of indirectly probing PeVatron particle distribution in the GC.

\begin{figure}
\begin{center}
\includegraphics[width=0.61\textwidth]{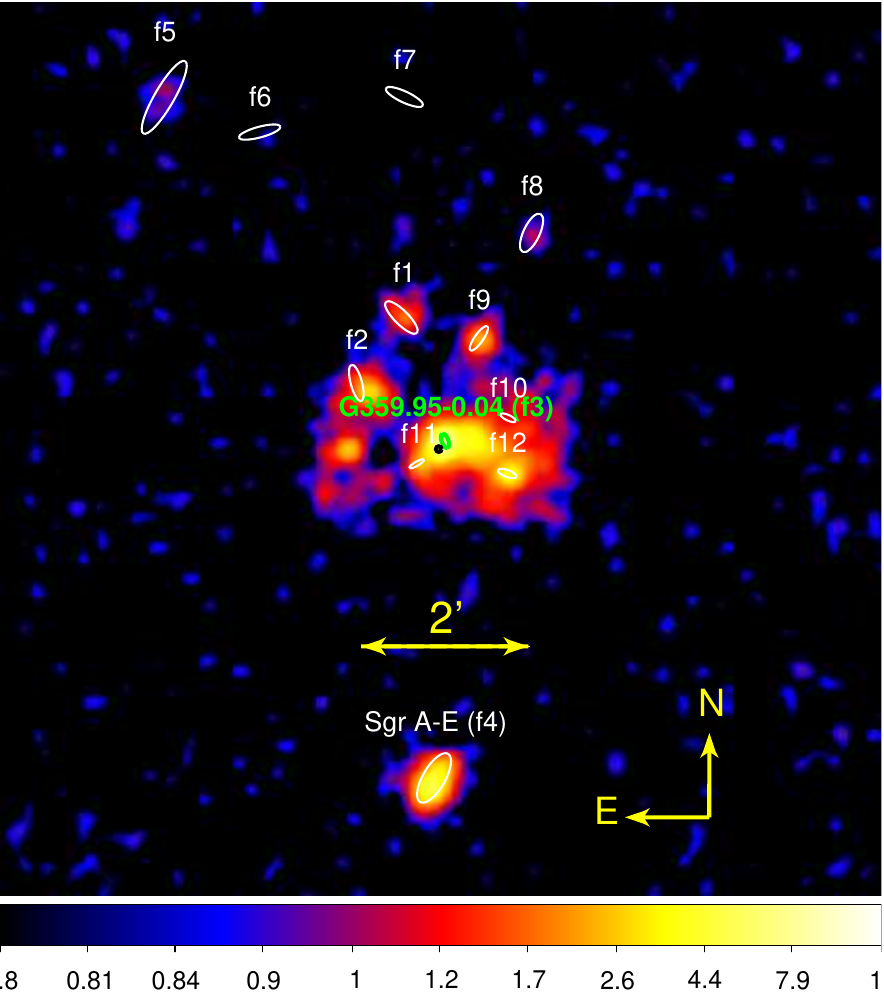}
\includegraphics[width=0.37\textwidth]{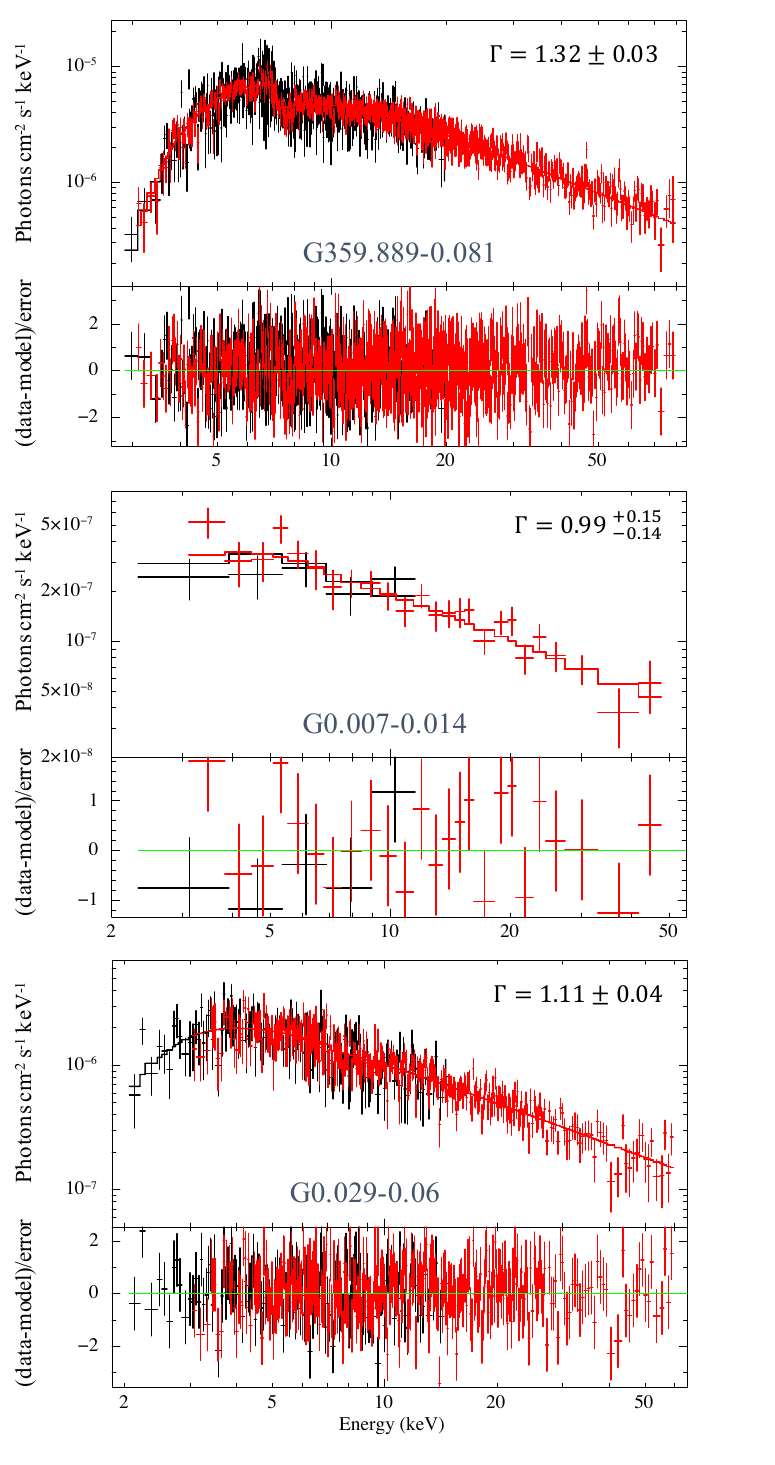}
\end{center}
\caption{\textit{Left:} Simulated \hexp\ 40--80 keV image of the GC region.  Eight of the 12 simulated non-thermal X-ray filaments are resolved. A list of the filaments and their spectral parameters is given in Table \ref{tab:filslist} in the Appendix. \textit{Right:} Simulated \hexp\ spectra of three of those filaments. LET and HET spectra are shown in black and red, respectively. \hexp\ will be able to constrain their spectral indices far better than current instruments.}\label{fig:het_gc_image}
\end{figure}

\hexp\ data above 40 keV will provide an unprecedentedly pristine view of non-thermal X-ray emission in the central 10 pc region. 
While the central PWN G359.95-0.04 and several X-ray filaments have been detected by \nustar\ above 40 keV \citep{Mori2015, Nynka2014NF}, other (potentially hidden) non-thermal X-ray sources remain largely unresolved. These non-thermal X-ray sources in the GC represent energetic particle accelerators (e.g., PWNe) or tracers of TeV--PeV electrons emitting synchrotron X-rays. 
It is still unclear just how high the maximum energy of X-rays emitted from the filaments 
around Sgr A* can reach \citep{Li2013}. This information is crucial for measuring the maximum energies of relativistic electrons in the region. 
\hexp\ is poised to resolve high-energy X-ray emission from a much larger number of X-ray filaments than \nustar\ and potentially discover new filaments. For example, a 500 ks {\hexp} simulation of the central $13' \times 13'$ field results in detecting 8 out of the 12 known X-ray filaments above 40 keV (Figure \ref{fig:het_gc_image}).  A full list of all filaments included in this simulation -- along with their spectral parameters -- is provided in Table \ref{tab:filslist} 
in the Appendix. 
Figure \ref{fig:het_gc_image} (right) shows simulated \hexp\ spectra for three filaments with 100~ks exposure: Sgr A-E (G359.889$-$0.081), G0.007$-$0.014 and G0.029$-$0.06, respectively. The simulated spectra can be fitted well with an absorbed power-law model extending beyond 50~keV. For brighter filaments like Sgr A-E and G0.029-0.06, 
spectral indices can be constrained to less than 5\%. A spectral break or cutoff below 50 ~keV can be ruled out. Determination of the maximum X-ray photon energy is essential to infer the energy of parent electrons. Unlike the current and proposed soft X-ray telescopes operating below 8--10 keV, broadband X-ray spectroscopy with \hexp\ can measure the X-ray spectral shapes robustly with minimal parameter degeneracy issues with $N_{\rm H}$. 
 Multi-wavelength SED and morphology studies of these filaments with MeerKAT/SKA and HEX-P will allow us to determine ambient B-field and broadband particle energy distribution. 
In summary, {\hexp} observation of X-ray and radio filaments will help to reveal the origin of relativistic particles in the GC, whether they are driven by common PeVatron accelerators in the GC (e.g., Sgr A* BH) or associated with local particle accelerators such as PWNe, SNRs, stellar winds or magnetic reconnections within the filaments.

\section{HEX-P science in other regions of the Milky Way} 

Besides the primary observation program of the GC and Bulge regions described above, below we list several unique \hexp\ science cases that can be explored through GO observations. These observation ideas have not been feasible with \nustar\ or stem from expanding the previous \nustar\ observations.

\subsection{Investigation of X-ray sources in star-forming regions \label{sec:star_forming}} 

Outside the GC and Bulge, a \hexp\ survey of several young massive clusters (YMCs) and star-formation regions in our Galaxy can be conducted to search for hidden BH--OB binaries, detect hard X-ray flares from young stellar objects (affecting the dynamics of protostellar disks) and identify hard X-ray sources previously detected by \nustar. YMCs, ranging in age from approximately 1 to 10 million years, are characterized by the presence of massive star-forming regions. These clusters offer valuable insights into ongoing star formation and often host compact objects such as NSs and stellar-mass BHs. This is due to the relatively short lifetimes of massive stars, which eventually undergo core-collapse supernovae. The Orion Nebula is particularly well-suited for studying a wide range of X-ray activities, including hard X-ray emission from BH binaries and flares from massive young stars. The Norma spiral arm, where \nustar\ detected $\sim30$ hard X-ray sources, could be surveyed with \hexp\ searching for HMXBs \citep{Fornasini2017}. 

\subsubsection{Search for black hole binaries with OB stars \label{sec:bh_ob}} 

It is estimated that the Milky Way harbors a substantial number of stellar mass BHs, ranging from $10^8$ to $10^9$. However, only around 20 BHs have been dynamically confirmed, and approximately 50 candidates have been identified mainly based on their X-ray characteristics. This discrepancy highlights the scarcity of confirmed BHs compared to the estimated population.  In addition, these confirmed BH systems are predominantly LMXBs, in contrast to the dearth of BH-HMXBs.
Currently, there are only six known HMXBs that contain a BH companion. Among them, Cygnus X-1 is the only system with an undisputed BH companion. Thus, the discovery of X-ray emission from just a few new BH--OB systems would significantly impact our understanding of these systems and their evolution.

The majority of massive stars are formed as binary or high multiple systems, as supported by studies such as \citet{Moe2017}. \citet{Langer2020} estimated that $\sim3$\% of massive OB stars have companions in the form of BHs, resulting in $\sim1,200$ BH--OB systems existing within our Galaxy. The discovery of these systems would provide valuable insights into various aspects of astrophysics, including star formation processes, stellar and binary evolution, galaxy evolution, supernova rates, and gravitational wave events. Through precise proper motion measurements conducted by the \gaia\ mission, two BH--OB systems named Gaia BH1 and BH2 have been recently discovered using their astrometric orbital solutions \citep{El-Badry2023a, El-Badry2023b}. These groundbreaking findings have opened new avenues of research in the field. It is anticipated that \gaia\ Data Release 3 will unveil $\sim190$ BH--OB binaries \citep{Janssens2022}, which may revolutionize our understanding of these systems. It is worth noting that \gaia\ would be sensitive to detecting the BH--OB systems with long orbital periods ($\gtrsim$ 10 days). These binaries are non-interacting in nature, similar to the initial two detections \citep[Gaia BH1 with an orbital period of 185.6 days and Gaia BH2 with 1277 days; see Figure 12 in][]{El-Badry2023b}. 

While \gaia's sensitivity does extend to systems with shorter orbital periods, the detection of BH--OB systems with periods $\lesssim$ 10 days can be facilitated by observing X-ray emission resulting from wind accretion onto the BH \citep{Gomez2021}. This is because BH--OB systems in close binaries with short orbital periods are more likely to have accretion onto the BH from the powerful stellar wind of the OB star, leading to the generation of X-ray emission. In particular, the presence of hard X-ray emission above 10 keV serves as a definitive signature of the BH's existence, especially when combined with knowledge of the orbital mass (which allows for differentiation between BH--OB and
NS--OB systems). This is due to the fact that single OB stars or (massive) active binaries rarely emit hard X-rays. Eta Carinae stands as the sole known massive binary system exhibiting hard X-rays above 10 keV. Thus, observing such hard X-ray emission can provide crucial evidence for the presence of a BH in an BH--OB system. 

In \citet{Gomez2021}, a novel method for identifying BH--OB systems is presented. The approach involves initially compiling a list of single-line spectroscopic binaries that exhibit no observable stellar spectrum of the secondary (with absorption lines) companion (hence, the binary is a SB1). Such systems are strong candidates to have secondary companions that are evolved stellar remnants: either a NS or a stellar-mass BH. These selected systems are then subjected to follow-up observations using \nustar\ to investigate the presence of hard X-ray emission originating from the binary. The critical signature of a NS or BH secondary is that the X-ray spectrum is detectable above 10 keV since isolated O stars have soft thermal spectra with $kT$ $\sim$ 1 - 3 keV and $L_X$ $\sim$ $10^{-7}$ $L_{bol}$ $\sim10^{31}$ erg s$^{-1}$. 
According to \citet{El-Badry2023a, El-Badry2023b}, BH--OB systems with $P_{\rm orb} \simlt 10$ days are expected to exhibit X-ray fluxes greater than $\sim$ $10^{-13}$ erg s$^{-1}$ cm$^{-2}$. These estimations suggest that both soft and hard X-ray emission from close BH--OB systems in nearby massive clusters are detectable with \hexp, for example, by surveying the central $23' \times 23'$ field of the Orion nebula, containing 1,600 \chandra\ X-ray sources \citep{Getman2005}, with 2 $\times$ 2 pointings (each with 60 ks exposure) and an additional 60 ksec observation at the center. \hexp's hard X-ray search for BH--OB binary candidates in several YMCs (alternatively, hard X-ray sources identified by \hexp\  in these fields can be further investigated using follow-up optical/infrared spectroscopic studies to measure orbital mass) will complement the BH systems discovered through astrometric orbital solutions with \gaia, which are primarily sensitive to non-interacting, long-period systems.

\subsubsection{Hard X-ray Flares from Young Stellar Objects \label{sec:yso}} 

Flares from stars are the most energetic events in their star systems, but our understanding of them, including flare distributions and their long-term influence on orbiting planets, is limited, primarily to our own sun. The situation becomes even more uncertain for flares from young stellar objects (YSOs), such as pre-main sequence stars (PMS). Deep X-ray surveys, like COUP, offer valuable opportunities to gain knowledge about flares originating from YSOs. According to the standard picture, a flare occurs due to magnetic reconnection, which accelerates non-thermal electrons along magnetic field lines, resulting in the heating of the chromosphere and corona. Based on the largest flares observed in COUP, \citet{Favata2005} suggested the existence of additional flare channels in YSOs, where reconnection events in magnetic field lines connect the star and the protoplanetary disk. \citet{Getman2008a, Getman2008b} also found that the brightest flare distribution, in terms of peak luminosity or temperatures, significantly deviates from expectations based on extrapolations from main sequence stars.

The evolution of a protoplanetary disk relies on the strength of the interaction between the magnetic field and ionized matter within the disk. High-temperature flares emit energetic photons that play a significant role in ionizing the disk, creating what is known as an active zone since they penetrate deep into the disk. The active zone, in turn, facilitates the transport of matter throughout the disk \citep{Glassgold2004} through mechanisms such as the magnetorotational instability \citep{Glassgold1997}. The influence of hard X-rays is particularly important as they penetrate deeper into the disk, making the hard X-ray luminosity a key factor in determining the accretion rate of a protostar.

\begin{figure}[h!]
\begin{center}
\hspace{-0.35cm}
\includegraphics[width=0.45\textwidth]{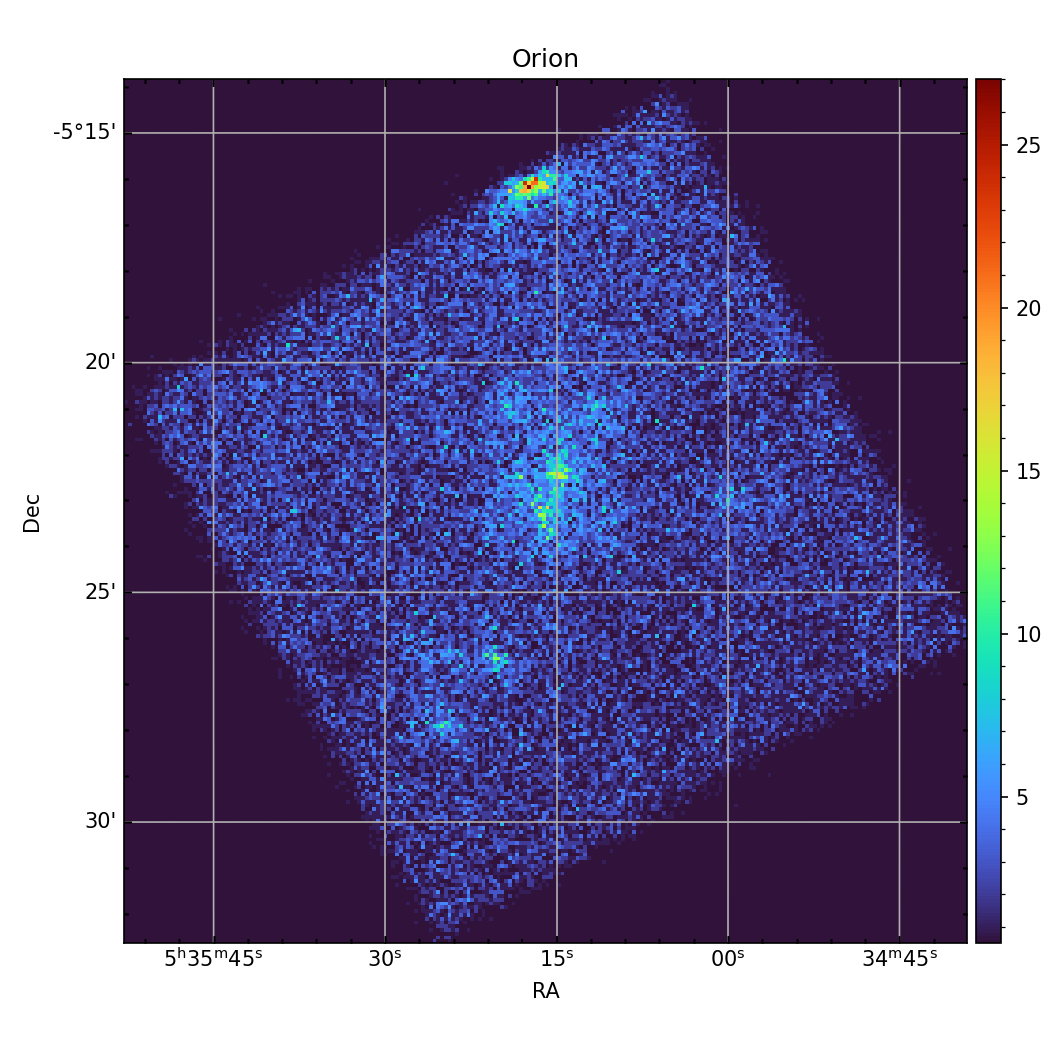}
\includegraphics[width=0.55\textwidth]{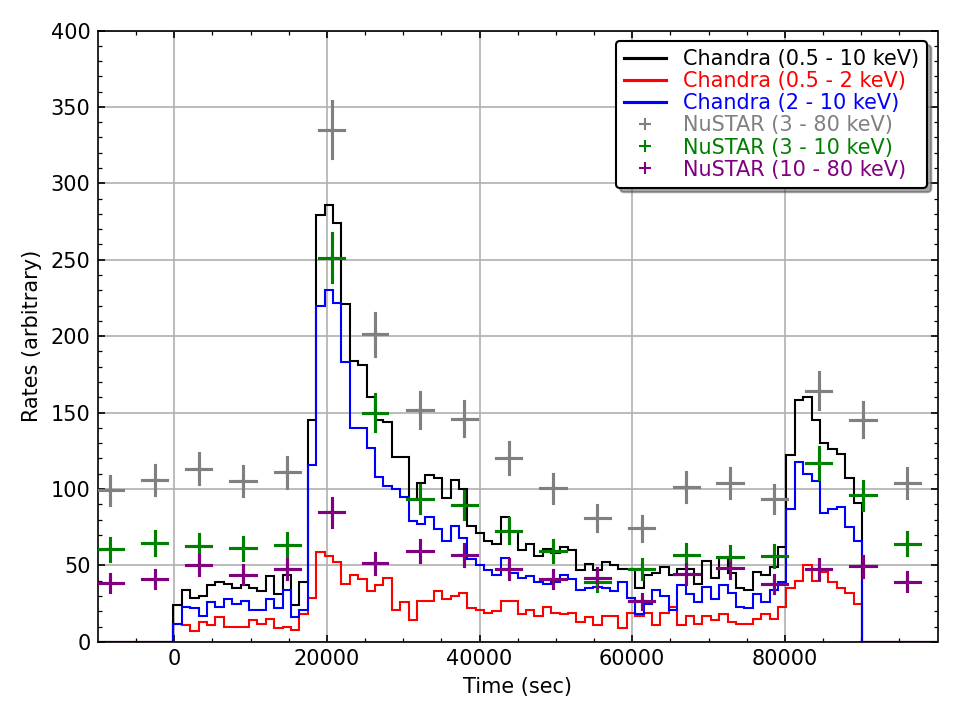}
\end{center}
\vspace{-0.5cm}
\caption{\nustar\ images of the Orion Nebula Cluster (left) and light curve of one of the 
variables (right). 
The hard ($>10$ keV) X-ray component (purple crosses) is heavily contaminated by neighboring sources, which makes hard X-ray flares difficult to distinguish with \nustar.  \hexp's superior angular resolution and sensitivity will enable us to identify many more flares from YSOs.}\label{fig:orion}
\end{figure}

The simultaneous observation of the Orion Nebula using \nustar\ and \chandra\ aimed to study the properties of flares from 
YSOs in the hard X-ray range ($>$10 keV) (Guenther 2023 in preparation). The \nustar\ data revealed a wide range of hard X-ray activity in the region, as depicted in  Figure~\ref{fig:orion}. However, the angular resolution of \nustar\ is relatively poor, leading to contamination from nearby sources in this densely populated area, which limits the hard X-ray investigation. Although $\sim8$ flares and $\sim6$  variables were detected, and some exhibited a similar pattern in their lightcurves between soft and hard X-rays, the \nustar\ lightcurves exhibited significant fluctuations due to low statistics and contamination from neighboring sources (Figure~\ref{fig:orion}). In contrast, due to its significantly improved sensitivities compared to \nustar, \hexp\ will enable a comprehensive study of five times more flares in both the soft and hard X-ray bands. Such a study will provide valuable insights into the evolution of protoplanetary disks and the formation of planets. 

\subsection{Magnetic Cataclysmic Variables \label{sec:mcv}}

The knowledge of WD masses, metallicities, and source types (e.g., magnetic vs. nmCVs) offers valuable insights into the formation and evolution of CVs in various environments within our Galaxy \citep{Mukai2017}.   
The high sensitivity and wide spectral coverage of \hexp\ present a unique opportunity to investigate the spatial evolution of the seemingly distinct CV populations in the GC, Bulge, Ridge and solar neighborhood. 
Also, as described in the next section, accurately measuring the WD masses of mCVs -- a task that \hexp\ is uniquely suited for -- allows us to probe the fundamental properties of these systems, such as (1) the long-term mass evolution of mCVs toward the Chandrasekhar mass limit with IPs, and (2) the origin of WD magnetic fields with polars. 
In the near future, \erosita, ZTF, and Rubin all-sky surveys are expected to discover a large number of mCVs in the Galactic Disk, including rare classes of CVs such as AM CVn stars, novae, and fast-spinning CVs \citep{Rodriguez2023}. \hexp\ will provide broad-band X-ray spectral and timing data of these exotic CVs, allowing us to study and understand those new populations with physically meaningful interpretations.

\subsubsection{White dwarf mass measurements from magnetic CVs}\label{WDmass}

Whether mCVs gain or lose mass is of great significance for probing 
the two competing channels for type Ia supernovae (SNIa) \citep{Pala2017}. 
In the double-degenerate case,
two WDs in a binary system merge 
to form a WD with $M > M_{Ch}$ (the Chandrasekhar mass limit, $M_{\rm WD} \sim 1.4 M_\odot$);  
in the alternative single-degenerate model, a
CO WD 
increases its mass at it 
accretes matter from its companion, eventually exceeding the Chandrasekhar limit.  
It is, however, still unknown whether CVs increase in mass over time due to accretion or lose mass through episodic nova eruptions. 
Interestingly, some of the newly identified IPs originally detected by \textit{INTEGRAL} were found to 
have WD masses 
close to the Chandrasekhar limit  
\citep{Tomsick2016, Tomsick2023}. 
However, the \textit{INTEGRAL} sources are intrinsically biased toward a population of more massive IPs, which  
generally possess harder X-ray spectra. In order to robustly test the CV mass evolution models such as the eCAML model \citep{Schreiber2016}, it is necessary to collate WD masses and characterize their distribution from a larger, unbiased sample of IPs. Polars, on the other hand, 
are more suited for studying the WD mass and B-field correlation, in order to test 
the common envelope (CE) model. The CE model is a leading but unconfirmed candidate for WD B-field formation where differential rotation between the decaying binary orbit and the  
CE leads to a magnetic dynamo, enhancing the B-field in the WD. One of the most robust predictions of CE models is an
anti-correlation between $B$ and $M_{\rm WD}$ \citep{Briggs2019}. Measuring the WD masses of polars will allow us to explore the $B$ vs $M$ correlation because their B-fields ($B = 7\rm{-}240$ MG) are well determined in the optical band. Since polars are typically fainter and more variable than IPs, \hexp's high sensitivity in the broad X-ray band will be crucial for determining their WD masses accurately.

\begin{figure}[h!]
\begin{center}
\includegraphics[width=0.5\textwidth]{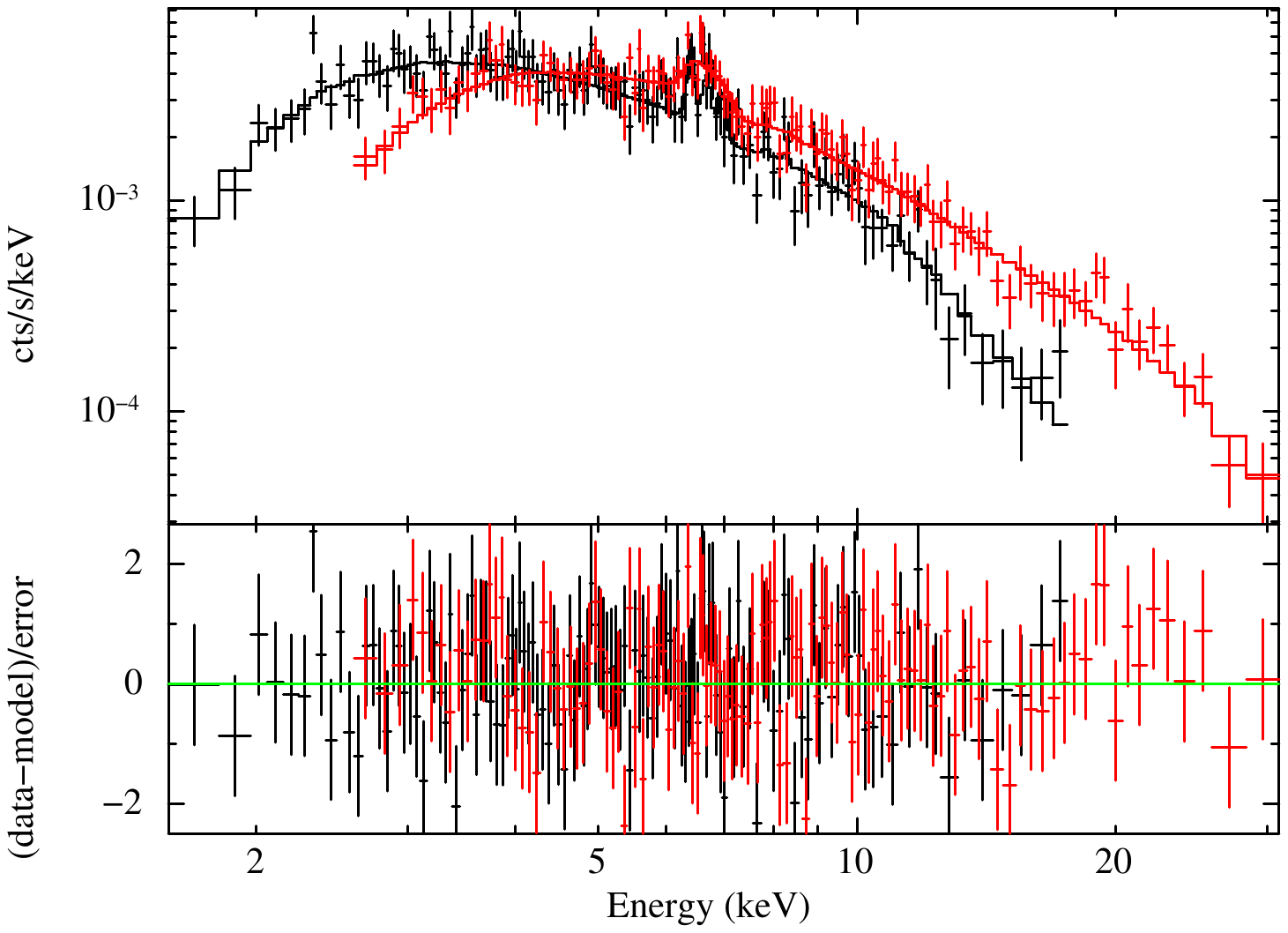}
\includegraphics[width=0.46\textwidth]{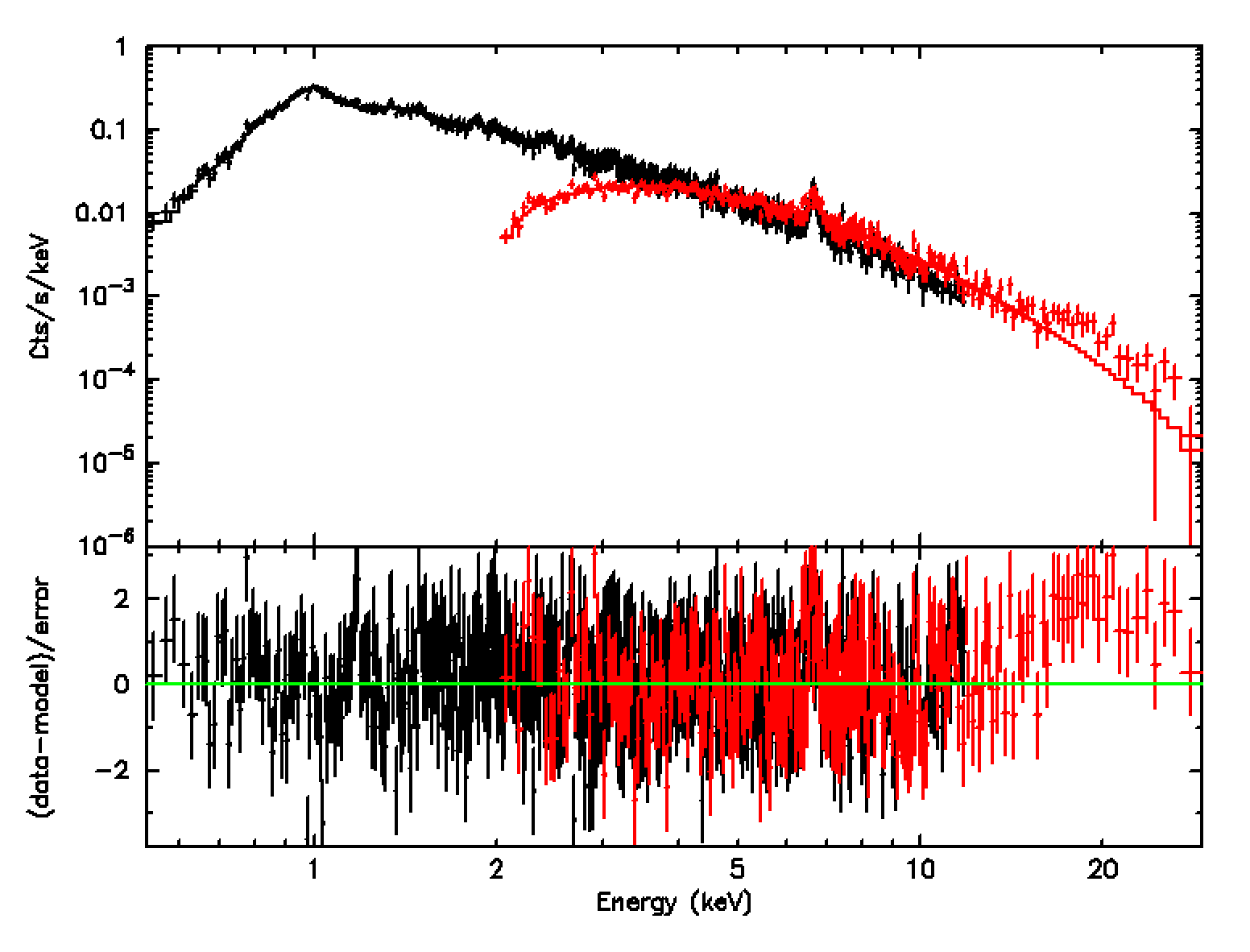} 
\end{center}
\caption{Left: Simulated \hexp\ LET (black) and HET (red) spectra of an IP with $M_{\rm WD} = 0.8 M_\odot$. Right: Simulated 100-ks \hexp\ LET (black) and HET (red) spectra of AE Aqr. The input model is composed of two thermal (APEC) and power-law components adopted from \citep{Kitaguchi2014}. The simulated \hexp\ spectra are fit only with the two thermal components to illustrate the hard non-thermal X-ray excess above 10 keV. }\label{fig:mcv_spectra}
\end{figure}

X-ray spectral modeling of the accretion flow  
over a broad energy range (extending beyond $\sim10$ keV)
is a particularly effective method for measuring WD masses
\citep{Shaw2020, Vermette2023}. In mCVs, infalling material is funneled onto the WD poles along B-field lines and heated to temperatures ($kT_{\rm shock}\sim10\rm{-}80$ keV) that scale with the WD mass. Below a standoff shock where the plasma temperature is highest, a column of infalling material cools via thermal bremsstrahlung and cyclotron radiation as it approaches the WD surface with varying temperatures and densities. Since hard X-rays are emitted from the post-shock region, hard X-ray data above $E\sim10$~keV provide the most accurate measurements of 
WD masses \citep{Hailey2016}. 
With the advent of more sophisticated X-ray spectral models for mCVs (e.g., \citet{Suleimanov2019, Hayashi2021}), \nustar's broadband X-ray spectral data led to measuring WD masses of $\sim30$ IPs \citep{Shaw2020}, including new IPs discovered in the Bulge accurately ($\simlt 20$\%) \citep{Mondal2022, Mondal2023}. \hexp\ will expand this program to measure WD masses from over 100 mCVs in the solar neighborhood ($d\simlt{\rm few}\times100$ pc), 
 allowing us to parse WD mass data into different types, regions, and accretion rates.  
Figure  
\ref{fig:mcv_spectra} (left) shows simulated \hexp\ spectra of an IP yielding WD mass measurements with 10\% error. 

\subsubsection{Exotic classes of magnetic CVs \label{sec:exotic_cvs}} 

Among the thousands of CVs discovered by extensive optical and X-ray surveys, several rare classes of CVs have been identified, such as AM CVn stars (compact CVs with short orbital periods), recurrent novae, and fast-spinning CVs. While rare, the populations and high-energy emission mechanisms of these exotic CVs are considered extremely important in contemporary astrophysics. For example, given their tight orbits, AM CVn stars are among the most likely gravitational wave sources detectable by the future \textit{LISA} observatory. 
In another example,
recurrent novae 
such as RS Oph, which undergo
episodic thermonuclear explosions on the WD surface 
that produce 
shocks from the ejecta slamming into the surrounding circumstellar winds, 
are of great interest for their
strong X-ray and gamma-ray emission \citep{Cheung2022}.  
Of particular and increasing interest is a rare class of fast-spinning CVs (FSCVs) with $P_{\rm spin} \simlt$ a few minutes. 
These FSCVs discovered so far are unique and distinct from regular accretion-powered CVs,  including the WD ``pulsar" systems (AR Sco and J191213.72-441045.1) and propeller CV (AE Aqr) \citep{Pelisoli2023}. 
In the near future, a larger population of undetected FSCVs may be revealed by the {\it Rubin} and \erosita\ all-sky surveys \citep{Pretorius2014}.  
\hexp's higher sensitivity and  
broad-band X-ray  
coverage will allow us to study a large number of FSCVs better than  \nustar. For example, \hexp\ will be able to detect the putative non-thermal X-ray component from AE Aqr above 10 keV (Figure \ref{fig:mcv_spectra} right panel). As demonstrated for AR Sco \citep{Takata2018}, multi-wavelength observations of new WD pulsars will be invaluable for understanding their particle acceleration and non-thermal emission mechanisms.

\section{Conclusions \label{sec:conclusion}} 

In conclusion, \hexp, as a probe-class mission, will offer an unprecedented opportunity to revolutionize our understanding of various astrophysical phenomena in the GC and broadly in Galactic astrophysics. With its high spatial resolution X-ray imaging, broad spectral coverage, and superior effective area, \hexp\ is poised to provide groundbreaking insights into a variety of important questions in the GC.  These include the past and current X-ray flares from the SMBH at Sgr A*, the populations of compact object binaries from the NSC to the GRXE, the primary particle accelerators, and cosmic-ray distributions in the GC. Outside the GC, \hexp\ is expected to accurately measure WD masses of hundreds of mCVs, explore broadband X-ray spectral and timing properties of X-ray transients (including determining BH spins from Galactic BH transients), detect pulsars, and search for BH--OB binaries in young star clusters. These scientific objectives can be uniquely achieved by \hexp\ or in synergy with other currently operating or future telescopes such as EHT, GRAVITY, MeerKAT, Roman, CTAO, and IceCube gen2. This paper, along with other Galactic science papers, highlights \hexp's enormous potential to uncover significant new insights into the most important astrophysical problems in the field of Galactic astrophysics in the 2030s.


\section*{Author Contributions}

\S1 (Mori). \S2 (Madsen and Garcia). \S3 (Mori, Mandel). \S4 (Stel, Ponti). \S5 (Mori, Mandel, Hong, Bachetti, Bodaghee, Bao, Ponti). \S6 (Kuznetsova, Mandel, Zhang, Krivonos, Clavel). \S7 (Hong, Grindlay, Mori, Rodriguez). \S8 (Mori).

\section*{Acknowledgments}
We are grateful to J. Wilms, T. Dauser, C. Kirsch, M. Lorenz, L. Dauner, and the SIXTE development team for their assistance with SIXTE simulations.

\bibliographystyle{Frontiers-Harvard} 
\bibliography{references,g1p9}

\clearpage
\appendix

\section*{Appendix}

\renewcommand{\arraystretch}{1.2}
\setlength{\tabcolsep}{30pt}
\begin{table*}[hb!]
\begin{center}
\begin{tabular}{ l l }
\hline \hline 
Acronym	&	Meaning\\
\hline
AGN     &   active galactic nucleus \\
BH      &   black hole \\
BH-LMXB &   black hole low-mass X-ray binary \\
CHXE    &   central hard X-ray emission \\
CMZ     &   Central Molecular Zone \\
CR      &   cosmic ray \\
CV      &   cataclysmic variable \\
FOV     &   field of view \\
FSCV    &   fast-spinning cataclysmic variable \\
GC      &   Galactic center \\
GRXE    &   Galactic ridge X-ray emission \\
HET     &   High Energy Telescope \\
HEX-P   &   High Energy X-ray Probe \\
HVS     &   hypervelocity star \\
IP      &   intermediate polar \\
IR      &   infrared \\
LECR    &   low energy cosmic rays \\
LET     &   Low Energy Telescope \\
LLIP    &   low-luminosity intermediate polar \\
LMXB    &   low-mass X-ray binary \\
LW      &   limiting window \\
MC      &   molecular cloud\\ 
mCV     &   magnetic CV \\
NIR     &   near infrared \\
NS      &   neutron star \\
NS-LMXB &   neutron star low-mass X-ray binary \\
NSC     &   nuclear star cluster \\
pc      &   parsec \\
PMS     &   pre-main sequence \\
PSF     &   point spread function \\
PWN     &   pulsar wind nebula \\
SED     &   spectral energy distribution \\
SMBH	&	supermassive black hole \\
SNIa    &   type Ia supernova \\
SNR     &   supernova remnant \\
ToO     &   target of opportunity \\
VFXT    &   very faint X-ray transient \\
WD      &   white dwarf \\
XRB     &   X-ray binary \\
YSO     &   young stellar object \\
\hline
\end{tabular}
\caption{List of acronyms used in this paper, along with their meanings. \label{tab:acronyms}}
\end{center}    
\end{table*}

\renewcommand{\arraystretch}{1.4}
\setlength{\tabcolsep}{18pt}
\begin{table*}
\begin{threeparttable}
\begin{center}
\caption{\hexp's sensitivity limits for detecting and characterizing X-ray sources in the GC}
\begin{tabular}{l >{\centering\arraybackslash}p{6cm} c}
\hline\hline
Goals & Measurement criteria & $L^{\rm min}_X$  [erg\,s$^{-1}$]\tnote{*}   \\ [1.25pt]
\hline\hline
Source classification &  Hardness ratio (3-10 vs 10-40 keV) with $<20$\% error &  $1\times10^{32}$ \\
Source identification &  Discerning between thermal and PL spectra with $>4\sigma$ significance  & $3\times10^{32}$  \\
Parameter determination & $kT$ and $\Gamma$ with 20\% error &  $8\times10^{32}$  \\ 
Variability detection &  X-ray flux with $<10$\% error &  $1\times10^{32}$   \\ 
Periodicity detection ($P \simgt 1$ hr) &  $> 4\sigma$ significance & $3\times10^{32}$    \\
Periodicity detection ($P \simlt 1$ sec) & $> 4\sigma$ significance & TBD  \\ 
\hline
\label{tab:point_sources}
\end{tabular}
\vspace{-0.25cm}
\begin{tablenotes}
    \item[*] The column labeled by $L^{\rm min}_X$ refers to the X-ray luminosity above which the measurement criteria will be met. We assumed 100 ks exposure per field for estimating the limiting $L^{\rm min}_X$ values based on XSPEC simulations. 
\end{tablenotes}
\end{center}
\end{threeparttable}
\end{table*}

\renewcommand{\arraystretch}{1.5}
\setlength{\tabcolsep}{14.5pt}
\begin{center}
\begin{table}[hb!] 
\caption{The filaments incorporated in the GC simulations described in Sections 5.4 and 8.2.  While best-fit values from current observations are very poorly constrained, \hexp\ will be able to measure their spectral properties with much better sensitivity (Section 8.2).  \label{tab:filslist}}
\begin{threeparttable}
\begin{tabular}{ l l c c r } 
 \hline\hline
 Label	&	Name	&	$N_{\rm H}$\tnote{a} 	&	$\Gamma$\tnote{a}	&	Flux (2-10 keV) \\
 & & [$10^{22} \, \rm{cm}^{-2}$] & & [erg cm$^{-2}$ s$^{-1}$] \\  [0.5ex]
 \hline
f1	&	G359.97-0.038	&	$11.7${\raisebox{0.5ex}{\tiny$^{+5.1}_{-1.9}$}}	&	$1.4${\raisebox{0.5ex}{\tiny$^{+0.8}_{-0.3}$}}	&	1.1$\times10^{-13}$ \\
f2	&	G359.964-0.052	&	$11.1${\raisebox{0.5ex}{\tiny$^{+2.4}_{-1.5}$}}	&	$1.9${\raisebox{0.5ex}{\tiny$^{+0.5}_{-0.3}$}}	&	2.3$\times10^{-13}$ \\
f3	&	G359.95-0.04\tnote{b}	&	$6.0${\raisebox{0.5ex}{\tiny$^{+2.0}_{-1.0}$}} ($12$) 	&	$1.8${\raisebox{0.5ex}{\tiny$^{+0.3}_{-0.2}$}}	&	6.0$\times10^{-13}$ \\
f4	&	G359.889-0.081 (Sgr A-E)	&	$31.4${\raisebox{0.5ex}{\tiny$^{+5.4}_{-2.0}$}}	&	$1.3${\raisebox{0.5ex}{\tiny$^{+0.6}_{-0.2}$}}	&	4.7$\times10^{-13}$ \\
f5	&	G0.029-0.06	&	$6.3${\raisebox{0.5ex}{\tiny$^{+1.9}_{-2.8}$}}	&	$1.1${\raisebox{0.5ex}{\tiny$^{+0.4}_{-0.4}$}}	&	1.2$\times10^{-13}$ \\
f6	&	G0.017-0.044	&	$0.0${\raisebox{0.5ex}{\tiny$^{+22.0}_{-0.0}$}} ($12$)	&	$-0.7${\raisebox{0.5ex}{\tiny$^{+3.0}_{-6.0}$}} ($2$)	&	4.0$\times10^{-14}$ \\
f7	&	G0.007-0.014	&	$5.7${\raisebox{0.5ex}{\tiny$^{+16.8}_{-5.7}$}}	&	$1${\raisebox{0.5ex}{\tiny$^{+3.4}_{-1.7}$}}	&	2.0$\times10^{-14}$ \\
f8	&	G359.97-0.009	&	$9.6${\raisebox{0.5ex}{\tiny$^{+8.8}_{-5.4}$}}	&	$1.2${\raisebox{0.5ex}{\tiny$^{+0.7}_{-0.6}$}}	&	4.0$\times10^{-14}$ \\
f9	&	G359.96-0.028	&	$7.2${\raisebox{0.5ex}{\tiny$^{+4.5}_{-3.2}$}}	&	$0.9${\raisebox{0.5ex}{\tiny$^{+0.6}_{-0.4}$}}	&	5.0$\times10^{-14}$ \\
f10	&	G359.942-0.03	&	$58.8${\raisebox{0.5ex}{\tiny$^{+32.8}_{-32.8}$}} ($12$)	&	$4.1${\raisebox{0.5ex}{\tiny$^{+3.0}_{-2.4}$}} ($2$)	&	2.0$\times10^{-14}$ \\
f11	&	G359.94-0.05	&	$-0.0${\raisebox{0.5ex}{\tiny$^{+9.3}_{-0.0}$}} ($12$)	&	$-0.4${\raisebox{0.5ex}{\tiny$^{+1.4}_{-1.0}$}} ($1$)	&	3.0$\times10^{-14}$ \\
f12	&	G359.933-0.037	&	$8.3${\raisebox{0.5ex}{\tiny$^{+5.8}_{-4.2}$}}	&	$0.7${\raisebox{0.5ex}{\tiny$^{+0.4}_{-0.5}$}}	&	5.0$\times10^{-14}$ \\  [0.5ex]
\hline
\end{tabular} 
\begin{tablenotes}
    \item [a] Best-fit parameters from \cite{Johnson2009} were used to model the filaments in our simulation, except for f3 (as described below), and for f6, f10, and f11, for which we used the values indicated in parenthesis.  This was done because of the exceptionally poor constraints and improbable best-fit values given for those filaments.  
    \item [b] The PWN G359.95-0.04 and its central pulsar were modeled separately for our simulations, assuming an absorbed power-law with $N_{\rm H}=12$, and $\Gamma=2.0$ and $1.5$, respectively.  These values are based on analysis by \cite{Wang2006}.
\end{tablenotes}
\end{threeparttable}
\end{table}
\end{center}

\end{document}